\documentclass[numberedappendix]{emulateapj}
\usepackage{epsf}
\begin{document}

\title{Evolution in the Halo Masses of Isolated Galaxies between
  $\lowercase{z}\sim1$ and $\lowercase{z}\sim0$: From DEEP2 to SDSS}

\author{
Charlie Conroy\altaffilmark{1}, 
Francisco Prada\altaffilmark{1},
Jeffrey A. Newman\altaffilmark{2},
Darren Croton\altaffilmark{3},
Alison L. Coil\altaffilmark{4},
Christopher J. Conselice\altaffilmark{5},
Michael C. Cooper\altaffilmark{3},
Marc Davis\altaffilmark{3,6},
S. M. Faber\altaffilmark{7},
Brian F. Gerke\altaffilmark{6},
Puragra Guhathakurta\altaffilmark{7},
Anatoly Klypin\altaffilmark{8},
David C. Koo\altaffilmark{7},
Renbin Yan\altaffilmark{3}
}

\altaffiltext{1}{Instituto de Astrofisica de Andalucia (CSIC), E-18008
  Granada, Spain}
\altaffiltext{2}{Hubble Fellow, Lawrence Berkeley National Laboratory, 
1 Cyclotron Road, Berkeley, CA 94720}
\altaffiltext{3}{Department of Astronomy, University of California,
Berkeley, CA 94720} 
\altaffiltext{4}{Hubble Fellow, Steward Observatory, University of Arizona,  
Tucson, AZ 85721}
\altaffiltext{5}{School of Physics and Astronomy, University of
  Nottingham, NG7 2RD, UK}
\altaffiltext{6}{Department of Physics, University of California,
Berkeley, CA 94720}
\altaffiltext{7}{University of California Observatories/Lick
Observatory, Department of Astronomy and Astrophysics, University of
California, Santa Cruz, CA 95064}
\altaffiltext{8}{Department of Astronomy, New Mexico State University,
  Box 30001, Department 4500, Las Cruces, NM 88003}

\begin{abstract}
  We measure the evolution in the virial mass-to-light ratio
  ($M_{200}/L_B$) and virial-to-stellar mass ratio ($M_{200}/M_\ast$)
  for isolated $\sim L^\ast$ galaxies between $z\sim1$ and $z\sim0$ by
  combining data from the DEEP2 Galaxy Redshift Survey and the Sloan
  Digital Sky Survey.  Utilizing the motions of satellite galaxies
  around isolated galaxies, we measure line-of-sight velocity
  dispersions and derive dark matter halo virial masses for these host
  galaxies.  At both epochs the velocity dispersion of satellites
  correlates with host galaxy stellar mass, $\sigma\propto
  M_\ast^{0.4\pm0.1}$, while the relation between satellite velocity
  dispersion and host galaxy $B$-band luminosity may grow somewhat
  shallower from $\sigma\propto L_B^{0.6\pm0.1}$ at $z\sim1$ to
  $\sigma\propto L_B^{0.4\pm0.1}$ at $z\sim0$.  The evolution in
  $M_{200}/M_\ast$ from $z\sim1$ to $z\sim0$ displays a bimodality
  insofar as host galaxies with stellar mass below $M_\ast
  \sim10^{11}$ $h^{-1} M_\Sun$ maintain a constant ratio (the
  intrinsic increase is constrained to a factor of $1.1\pm0.7$) while
  host galaxies above $M_\ast \sim10^{11}$ $h^{-1} M_\Sun$ experience
  a factor of $4\pm3$ increase in their virial-to-stellar mass ratio.
  This result can be easily understood if galaxies below this stellar
  mass scale continue to form stars while star formation in galaxies
  above this scale is quenched and the dark matter halos of galaxies
  both above and below this scale grow in accordance with $\Lambda$CDM
  cosmological simulations.  Host galaxies that are red in $U-B$ color
  have larger satellite dispersions and hence reside on average in
  more massive halos than blue galaxies at both $z\sim1$ and $z\sim0$.
  The satellite population of host galaxies varies little between
  these epochs; the only significant difference is that satellites at
  $z\sim1$ tend to be comparatively fainter (by $\sim0.15$ magnitudes
  in the mean) relative to their host luminosity than satellites at
  $z\sim0$.  The redshift and host galaxy stellar mass dependence of
  $M_{200}/M_\ast$ agrees qualitatively with the Millennium Run
  semi-analytic model of galaxy formation.
\end{abstract}


\section{Introduction}

The current cosmological framework indicates that galaxies are
embedded in massive dark matter halos that extend far beyond the
visible baryonic component.  The virial mass-to-light, $M_{200}/L$,
and virial-to-stellar mass, $M_{200}/M_\ast$, ratios provide simple
distillations of the complex interplay between galaxies and their dark
matter halos.  Constraints on the dependence of $M_{200}/L$ and
$M_{200}/M_\ast$ on galaxy properties and redshift hence afford unique
insight into the formation and evolution of galaxies in a cosmological
context.  For example, evolution of $M_{200}/M_\ast$ provides
constraints on the evolution of star formation efficiency in galaxies.

Yet there are surprisingly few direct constraints on $M_{200}/L$ or
$M_{200}/M_\ast$.  Since dark matter extends far beyond the visible
components of a galaxy, it is in practice difficult to probe the halo
mass on large scales (i.e., $\gtrsim100$ $h^{-1}$ kpc) due to a lack
of luminous tracers.  The halo masses of clusters of galaxies can be
estimated by strong and weak gravitational lensing, the tight
relationship between the X-ray temperature of the diffuse intracluster
medium and the dark matter halo, the Sunyaev-Zeldovich effect, and by
measuring the velocity dispersion of the cluster galaxies themselves,
under the assumption that the cluster galaxies are tracing out the
dark matter halo potential \citep[see][for a recent review]{Voit05}.

Probing the halos of isolated galaxies is more difficult. The only
methods that are currently capable of directly probing the halo mass
of isolated $\sim L^\ast$ galaxies to large radii are weak lensing
\citep{Brainerd96,Wilson01, Guzik02, Hoekstra04, Hoekstra05,
  Kleinheinrich05, Mandelbaum06} and satellite dynamics \citep{LT87,
  EGH87, EGH99,Zaritsky93,Zaritsky97, McKay02, Prada03,Brainerd03,
  Brainerd05, VDB04b,Conroy05a}, which utilizes satellite galaxies as
test particles that trace out the dark matter halo velocity field out
to several hundred kiloparsecs.  One limitation of both techniques is
that they must ``stack'' many isolated galaxies in order to build up a
robust signal.  Both methods have yielded similar results; e.g., each
finds a positive correlation between galaxy luminosity and the virial
mass of its dark matter halo.

The evolution in the virial mass-to-light ratios of bright isolated
galaxies is only poorly constrained.  \citet{Wilson01} measured the
weak lensing signal for early-type $\sim L^\ast$ galaxies from $z=0.8$
to $z=0.1$.  They found little evolution in the halo mass of $\sim
L^\ast$ galaxies, though their formal errors on the halo mass at
$z\sim0.8$ were $\gtrsim50$\%.  Furthermore, that work assumed that
$L^\ast$ did not evolve with redshift and that halo mass was
proportional to the square root of galaxy luminosity.  Most recently,
utilizing stellar masses and weak lensing data from COMBO-17 and GEMS,
\citet{Heymans06} find no significant evolution in the
virial-to-stellar mass ratios of bright galaxies to $z\sim0.8$ (though
constraints are weak; they find that the ratio decreases by no more
than a factor of $2.6$ at $1\sigma$).

Using the first $\sim25$\% of the recently completed DEEP2 redshift
survey, which has now collected spectra for $>40,000$ galaxies at
$0.7<z<1.4$, \citet{Conroy05a} used the dynamics of satellite galaxies
to measure the halo mass for bright isolated galaxies with satellites
(referred to as ``host'' galaxies) and compared their derived $M/L$ to
measurements from the Sloan Digital Sky Survey (SDSS) at $z\sim0$.
However, the small sample size and differences in selection effects
between DEEP2 and SDSS meant that all claims in that work had to be
highly qualified.

This study presents a much more detailed comparison between
host-satellite systems identified in the completed DEEP2 Redshift
Survey and systems found in a consistent way from the fourth public
data release of the SDSS.  The increased data sample at $z\sim1$,
combined with a careful handling of the different selection effects
between the two surveys, allows a robust determination of the
evolution in the $B$-band virial mass-to-light ratio and the
virial-to-stellar mass ratio of $\sim L^\ast$ host galaxies from
$z\sim 1$ to $z\sim 0$.

The classic Tully-Fisher relation is also capable of constraining the
halo mass of isolated disk-dominated galaxies, though the steps
required to convert the observed rotation speed into a dark matter
halo virial mass is more model dependent than either weak lensing or
satellite dynamics.  \citep{VDB02,Kassin06a, Gnedin06}.  In
particular, this method must extrapolate the rotation curve far beyond
the region covered by observations and/or requires knowledge of the
relative contribution of baryonic and dark matter to the observed
rotation curve as a function of scale \citep[see, e.g.,][]{Gnedin06}.
Recently, \citet{Conselice05}, \citet{Boehm06}, and \citet{Kassin06b}
have used the Tully-Fisher relation to constrain the evolution in the
virial-to-stellar mass ratio from $z\sim1$ to $z\sim0$.  These studies
found no evidence for a change in this ratio, though their sample
sizes were relatively small compared to weak lensing and satellite
kinematics studies ($\sim 100$ objects for the first two studies and
$\sim550$ for the last).

This article proceeds in the following manner: $\S$\ref{s:method}
outlines our definition for host and satellite galaxies and describes
our method for estimating halo masses of host galaxies.  In
$\S$\ref{s:data} we describe the galaxy surveys studied and explain
the methods used to mitigate survey selection effects.
$\S$\ref{s:props} contains a comparison of the properties of host and
satellite galaxies between $z\sim1$ and $z\sim0$, and in
$\S$\ref{s:results} we present satellite velocity dispersions and
derived virial mass-to-light and virial-to-stellar mass ratios as a
function of host galaxy redshift, luminosity, stellar mass, and color.
$\S$\ref{s:sam} contains a comparison between our results and a
semi-analytic model of galaxy evolution.  In $\S$\ref{s:disc} we
discuss these results and $\S$\ref{s:conc} concludes.  Those readers
not interested in the technical details should focus on $\S\S$
\ref{s:props} -- \ref{s:conc}.

A $\Lambda$CDM cosmology is adopted throughout:
$\Omega_m=1-\Omega_\Lambda=0.3$, with $H_0 = 100h$ km s$^{-1}$
Mpc$^{-1}$.  All absolute magnitudes quoted here are in the $AB$
system \citep{Oke83}, and throughout magnitudes are quote as
$M_B-5\rm{log}$$(h)$.  We adopt a mass definition for dark matter
halos such that the virial radius of a halo corresponds to a region
with mean density $200$ times the critical density, denoted $r_{200}$.
The halo mass is the mass interior to $r_{200}$, and is denoted
$M_{200}$.  $M_\ast$ is reserved for stellar masses and $M^\ast$ for
the characteristic scale of the luminosity function.

\section{Methodology}\label{s:method}
This section describes our definitions of host and satellite galaxies
and our methods for determining line-of-sight velocity dispersions and
dark matter halo masses of host galaxies.

\subsection{Isolation and Satellite Criteria}\label{s:params}

The use of satellite dynamics for extracting mass estimates of their
host galaxies is motivated by a scenario in which a bright galaxy
resides at rest at the center of its dark matter halo, has no other
bright companions, and is surrounded by faint satellites that are
bound to the host halo and orbit within it.  The criteria
used to define host galaxies and their associated satellites are meant
to capture such systems.  Note that, according to our use, ``host''
galaxies are not completely isolated; rather they are isolated with
respect to other comparably bright galaxies.  These criteria reject
objects in dense regions such as clusters and groups, which typically
contain several comparably bright galaxies.

Isolation criteria is specified using three parameters: an absolute
$B$-band magnitude difference, $\Delta M_B$, a velocity difference,
$\Delta V$, and a projected physical separation $\Delta R_p$.  The
latter two parameters define a search cylinder: if a galaxy has no
companions within the search cylinder that are within $\Delta M_B$ in
absolute magnitude, then it is deemed to be ``isolated''.

With a set of isolated galaxies we then search for satellite
companions by specifying a set of three similar parameters: a
magnitude difference, $\delta M_B$, a velocity difference, $\delta V$,
and a projected physical separation, $\delta R_p$ (here and throughout
we reserve $\Delta$ for isolation criteria and $\delta$ for satellite
criteria).  We then search for companions of the isolated galaxies
that are within the search cylinder defined by $\delta V$ and $\delta
R_p$, and are fainter than the isolated galaxy by at least $\delta M_B$
magnitudes.  Isolated galaxies with satellites are called ``host
galaxies''.

\begin{deluxetable}{llllllll}
\tablecaption{Search Criteria} \tablehead{
  \multicolumn{1}{c}{} &
  \multicolumn{3}{c}{\underbar{\hbox to 90pt{\hfill Isolation Criteria
\hfill}}}  & 
  \multicolumn{3}{c}{\underbar{\hbox to 90pt{
\hfill Satellite Criteria \hfill}}} \\
  \\
  \colhead{Sample} &
  \colhead{$\Delta M_B$} &
  \colhead{$\Delta V$} &
  \colhead{$\Delta R_p$} &
  \colhead{$\delta M_B$} & 
  \colhead{$\delta V$} &
  \colhead{$\delta R_p$} \\
  \colhead{Name} &
  \colhead{} &
  \colhead{km s$^{-1}$} &
  \colhead{$h^{-1}$ kpc} &
  \colhead{} & 
  \colhead{km s$^{-1}$} &
  \colhead{$h^{-1}$ kpc} }
\startdata
$A$ & 1.0 & 1000 & 500 & 1.0 & 750 & 350 \\
$B$ & 1.5 & 1000 & 500 & 1.5 & 750 & 350 \\
$C$ & 1.5 & 1000 & 1000 & 1.5 & 500 & 350 \\
\enddata
\tablecomments{The $\Delta V$ and $\delta V$ parameters are set to
  larger values when considering samples of brighter host galaxies.
  See $\S$\ref{s:params} for details. \vspace{0.5cm}}
\label{t:params}
\end{deluxetable} 

Various authors have used different parameters for identifying
host-satellite systems and have found that the resulting derived halo
mass is quite insensitive to reasonable choices of parameters
\citep{Prada03, Conroy05a}.  In this work we use the search parameters
listed in Table \ref{t:params}; the same parameters are used to
extract systems from both the DEEP2 and SDSS surveys, and our fiducial
set of parameters is sample $A$.  In Appendix \ref{s:test_mass} we
show that our results are unchanged within $1\sigma$ when adopting
different search parameters.

However, the recovered satellite velocity dispersion (see below) can
become sensitive to $\delta V$ (the maximum velocity separation
between host and satellite) when the contribution from the true
satellite dispersion is significant even at the edge of the $dV$
distribution (the window defined by $\pm\delta V$).  This problem is
alleviated simply by increasing the $\delta V$ parameter for brighter
host galaxies (which have larger satellite dispersions).  Indeed,
tests with simulations have shown that scaling the $\delta V$
parameter with host galaxy luminosity more robustly recovers the true
satellite dispersion at large host luminosities \citep{VDB04b,
  Chen06}.  Hence for the highest host galaxy luminosity and stellar
mass bins at $z\sim0$ and $z\sim1$, we increase $\delta V$ by
increments of $500$ km s$^{-1}$ until the measured dispersion has
converged (the satellite dispersion for fainter samples had already
converged using sample $A$ parameters).  Convergence is achieved when
using $\delta V=1000$ km s$^{-1}$ for all of these bins except for the
highest stellar mass bin at $z\sim0$; there the dispersion converges at
$\delta V=1500$ km s$^{-1}$.\footnote{In addition, for this case
  $\Delta V$ has been increased to $1500$ km s$^{-1}$ since it is
  conceptually awkward for $\delta V$ to be larger than $\Delta V$.}

\subsection{Velocity Dispersion \& Halo Mass Estimation}\label{s:mass_est}

This section describes in detail how to estimate the dark matter halo
mass of host galaxies.  This procedure can be conceptually separated
into three steps: identifying host-satellite systems in a galaxy
catalog; reconstructing the line-of-sight velocity dispersion profile
of their satellites; and converting the dispersion profile into a
mass.  Once host-satellite systems are found, the velocity dispersion
profile\footnote{Note that what is actually being probed is the
  \emph{line-of-sight} velocity dispersion profile; we drop
  `line-of-sight' in the remainder of this paper for brevity.}  of the
host galaxy dark matter halo can be estimated from the distribution of
velocity differences between host and satellite galaxies, $dV\equiv
V_{\rm{host}}-V_{\rm{sat}}$, in bins of projected distance from the
host galaxy, $R_p$.

In an ideal world there would exist hundreds of sufficiently luminous
satellite galaxies per host galaxy, and the task of measuring a
velocity dispersion profile would be comparatively straightforward.
Unfortunately, in practice there are on average only $1-2$ satellites
per host galaxy, due to the magnitude limit of the redshift surveys
used here (most isolated galaxies possess \emph{no} identifiable
satellites).  In order to build up a dispersion profile we must
combine satellites from many host galaxies and construct an average
profile around an average host \citep[e.g.][]{Zaritsky94}.  Assuming
that host galaxy properties such as luminosity are tightly correlated
with their dark matter halo mass, then by stacking host galaxies in
bins of luminosity we can recover the average underlying halo mass
\citep{Prada03}.  Weak lensing studies rely on the same stacking
procedure, since the lensing signal from individual isolated galaxies
is very weak \citep[e.g.,][]{Brainerd96}.

Naively, one might expect the velocity distribution of satellites,
$f(dV)$, to be approximately Gaussian with $\sigma$ given by the
velocity dispersion of satellite galaxies associated with the host
galaxy \citep{Prada03}.  In fact, there is a significant contribution
to the $dV$ distribution from so-called ``interloper'' galaxies.
Interlopers are galaxies that are classified as satellites in
projection but are in fact not true satellites, i.e., they are not
physically associated with the host galaxy.

Previous authors \citep[e.g.,][]{McKay02, Prada03, Conroy05a} have
modeled the effect of interlopers as a constant contribution to
$f(dV)$ at all velocities.  Tests with simulations have confirmed that
modeling interlopers in this way results in a robust recovery of the
underlying mass distribution (see Appendix \ref{s:appraisal}).  However,
\citet{VDB04b} and \citet{Chen06} found that in mock galaxy catalogs
interlopers do not have a constant $dV$ distribution but rather have a
velocity structure somewhat similar to true satellites (though in the
mock catalogs the width of the interloper distribution does not appear
to scale with host galaxy luminosity).  Despite this (or perhaps
\emph{because} of this), these authors found that modeling the
interloper distribution as a constant component to the $dV$
distribution does yield an accurate recovery of the velocity
dispersion profile as well as the underlying halo mass, in agreement
with previous work.

Hence $f(dV)$ is modeled as a Gaussian distribution plus a constant
component:
\begin{equation}
  f(dV; \, \eta,\sigma_{\rm{los}}) = \frac{\eta}{2\delta V} + 
  \frac{1-\eta}{\sqrt{2\pi}\,\sigma \, \mathrm{erf}(\delta V/\sqrt{2}\sigma)}\, 
  e^{-dV^2/(2\sigma_{\rm{los}}^2)},
\end{equation}
subject to the following normalization condition:
\begin{equation}
  \int^{+\delta V}_{-\delta V} f(dV; \, \eta,\sigma_{\rm{los}})  
  \, \mathrm{d}(dV) = 1,
\end{equation}
where $\eta$ is the fraction of interlopers within $\pm \delta V$ and
$\sigma_{\rm{los}}$ is the line--of--sight satellite dispersion.

We use a maximum--likelihood method to fit this
Gaussian--plus--constant function to the observed distribution of $dV$
for pairs in some bin of projected separation $R_p$.  We maximize the
likelihood function:
\begin{equation}
  \mathrm{ln}\big[L(\eta,\sigma_{\rm{los}})\big] = 
  \sum_i \lambda_i\, \mathrm{ln}\big[f(dV_i; \, \eta, \sigma_{\rm{los}})\big]
\end{equation}
over a dense grid in $\sigma_{\rm{los}}$ and $\eta$, where $dV_i$ is
the $dV$ value for the $i$th satellite--host galaxy pair, which is
given weight $\lambda_i$ (see below).  Since we only use $dV_i$
values for pairs in some bin of projected separation, the parameters
$\eta$ and $\sigma_{\rm{los}}$ are implicit functions of $R_p$.

In the following analysis satellite $i$ is assigned a weight
$\lambda_i$ according to the inverse number of satellites associated
with the host of satellite $i$.  This weighting scheme ensures that
host galaxies that have a large number of satellites (and are hence
likely to be more massive than the average host) do not dominate the
likelihood.  Van den Bosch et al. (2004) compared weighting by host
galaxies (the scheme employed here) to weighting by satellite galaxies
(which would be equivalent to setting $\lambda=1$ for all satellites)
in simulations and found that the recovered velocity dispersion
differs between these two schemes as a function of host
luminosity\footnote{This is due to the fact that brighter galaxies
  will have more satellites, especially in a flux limited sample, and
  hence the difference in weighting schemes is more pronounced in the
  regime of bright host luminosity.}.  Weighting by host galaxy more
fairly represents the average mass of host galaxies within a given
luminosity or stellar mass bin.

The resulting fit yields not only a measurement of the velocity
dispersion but also an estimate of the interloper fraction.
Marginalizing the likelihood over $\eta$ provides an estimate of the
$1\sigma$ errors on the velocity dispersion.

Redshift uncertainties are accounted for by subtracting in quadrature
the rms redshift error, $\sigma_{\rm{err}}$, from the measured
velocity dispersion, $\sigma_{\rm{los}}$:
\begin{equation}
\sigma_{\rm{est}} = \sqrt{\sigma_{\rm{los}}^2 - 2\, \sigma_{\rm{err}}^2}.
\end{equation}
The redshift uncertainty enters twice because we are subtracting two
velocities, the host from the satellite.  The resulting velocity
dispersion, $\sigma_{\rm{est}}$, is then our best estimate of the true
dispersion.  For the range of velocity dispersions probed here,
folding redshift errors from the DEEP2 and SDSS surveys
($\sigma_{\rm{err}}\lesssim30$ km s$^{-1}$) into our measured dispersion
changes results by only a few percent.  In the following sections we
simplify our notation to $\sigma=\sigma_{\rm{est}}$ for brevity.

With an estimate of the velocity dispersion of the host galaxy dark
matter halo we can, with additional assumptions, extract the virial
mass of the halo.  We follow the procedure of \citet{Prada03} and
\citet{Conroy05a} in converting velocity dispersions to virial masses.

The density profile of a dark matter halo is parameterized using the
NFW \citep{NFW97} model:
\begin{equation}\label{eqn:nfw}
\frac{\rho(r)}{\rho^0_c} = \frac{\delta_c}{(r/r_s)(1+r/r_s)^2}
\end{equation}
where $\rho^0_c$ is the present critical density, 
$r_s = r_{200}/c$, and 
\begin{equation}
\delta_c = \frac{200}{3} \frac{c^3}{\rm{ln}(1+c) - c/(1+c)},
\end{equation}
where $r_{200}$ is defined as the radius where the mean interior
density is $200$ times the critical density.  The concentration, $c$,
is inversely related to the mass of a dark matter halo, and, at fixed
mass, scales with redshift as $(1+z)^{-1}$ \citep{Bullock01};
concentrations for dark matter halos hosting galaxies or clusters
range from $3\lesssim c \lesssim 25$.  In our analysis we fix $c=10$,
which is consistent with the concentration of a $\sim10^{12}$ $h^{-1}
M_\Sun$ halo at $z=0$, and scale $c$ by $(1+z)^{-1}$ for higher
redshift samples.  However, as demonstrated below, the density profile
depends only weakly on concentration over the scales probed
($20\lesssim R_p\lesssim 150$ $h^{-1}$ kpc), and hence the assumed
concentration has little effect on the resulting mass estimates
(cf. Appendix \ref{s:test_mass}).  Throughout we assume that the
satellite galaxies follow the radial profile of the dark matter.  The
effect of a spatial bias between satellites and dark matter on the
recovered mass have been show by \citet{VDB04b} to be at the few
percent level (see also Appendix \ref{s:appraisal}).

We then determine the radial velocity dispersion profile by
integrating the Jeans equation, which relates the density profile and
gravitational potential to the radial velocity dispersion, using
Equation \ref{eqn:nfw} to specify the potential.  Finally, we
integrate along the line-of-sight.  Both integrations require
knowledge of the velocity anisotropy, $\beta \equiv
1-\sigma_{r}^2/\sigma_{\perp}^2$, of the satellite population.
Fortunately the line-of-sight velocity dispersion profile depends only
weakly on $\beta$ \citep{VDB04b,Mamon05}.  In Appendix
\ref{s:test_mass} we explore both an isotropic distribution
($\beta=0$) and an anisotropy parameterization suggested by
\citet{Mamon05}, and confirm that the derived mass is robust to
assumptions about $\beta$.  Below we set $\beta=0$.

Given these assumptions, there remains only one free parameter, the
normalization of the dispersion profile, which is related to the mass
within $r_{200}$, denoted $M_{200}$, of the dark matter halo.  The
normalization is obtained via $\chi^2$ minimization using the measured
$\sigma(R_p)$ points.  The majority of the analysis below uses
only one velocity dispersion measurement to derive a virial mass.  In
Appendix \ref{s:test_mass} we demonstrate that estimating virial masses
with only one dispersion measurement does not bias the recovered mass.
Including more velocity dispersion measurements in finer projected
separation bins simply has the effect of decreasing the error
on the recovered mass (if there are sufficient numbers of satellites
to increase the number of radial bins, which is not the case in
DEEP2).  See \citet{Klypin06} for a detailed study of the full
satellite velocity dispersion profile measured for SDSS host galaxies
at $z\sim0$.

\section{The Data}\label{s:data}
This section presents the low- and high-redshift galaxy catalogs
used to identify host and satellite galaxies, and describes how to
account for the differences in selection effects between the two
catalogs.

\subsection{The SDSS}
The Sloan Digital Sky Survey \citep[SDSS;][]{York00,Abazajian04,DR4}
is an extensive photometric and spectroscopic survey of the local
Universe.  As of Data Release 4, imaging data exist over $6670$
deg$^2$ in five bandpasses, $u$, $g$, $r$, $i$, and $z$.
Approximately $670,000$ objects over $4780$ deg$^2$ have have been
targeted for follow-up spectroscopy as part of SDSS are included in
DR4; most spectroscopic targets are brighter than $r=17.77$
\citep{Strauss02}.  Automated software performs all the necessary data
reduction including the assignment of redshifts.  Redshift errors are
$\lesssim 30$ km s$^{-1}$, similar to DEEP2.  The spectrograph tiling
algorithm ensures nearly complete sampling \citep{Blanton03a}, yet the
survey is not $100$\% complete due to several effects: 1) fiber
collisions do not allow objects separated by $<1'$ to be
simultaneously targeted, resulting in $\sim6$\% of targetable objects
failing to be targeted for spectroscopy; 2) a small fraction ($<1$\%)
of targeted galaxies fail to yield a reliable redshift; and 3) bright
Galactic stars block small regions of the sky.  None of these effects
is expected to impact our analysis.  The overall completeness of the
SDSS, as defined by the number of objects with successful redshifts
divided by the number of objects in the imaging catalog with
$r<17.77$, is $\sim90$\%.  The parent catalog has $166,923$ high
quality redshifts between $0.01<z<0.1$ and $100^o<RA<275^o$.

For this analysis we make use of the hybrid NYU Value Added Galaxy
Catalog (VAGC) \citep{Blanton05}.  In addition we use the publicly
available package \texttt{kcorrect v4.1.4} \citep{Blanton03b,
  Blanton06a} to derive restframe $B$-band magnitudes and $U-B$ colors
for SDSS galaxies.  All SDSS galaxies are K-corrected to $z=0.0$.
Galaxies are divided into red and blue populations based on the valley
visible in the color-magnitude diagram \citep[e.g.,][]{Baldry04}.  We
account for the fact that the valley moves redward for brighter
galaxies with the following color-cut:
\begin{equation}
U-B = -0.066 \,M_B - 0.05.
\end{equation}

We also obtain stellar masses for SDSS galaxies using \texttt{kcorrect
  v4.1.4} routines.  These stellar masses, which have been obtained
assuming a Chabrier IMF, are consistent with the
stellar masses of \citet{Kauffmann03a} but are lower by $\sim0.3$
dex from the color-based stellar mass estimates of \citet{Bell03}.
The offset is due primarily to differences in the assumed IMF.

\subsection{The DEEP2 Survey}\label{s:deep2}

\begin{figure}
\centering
\plotone{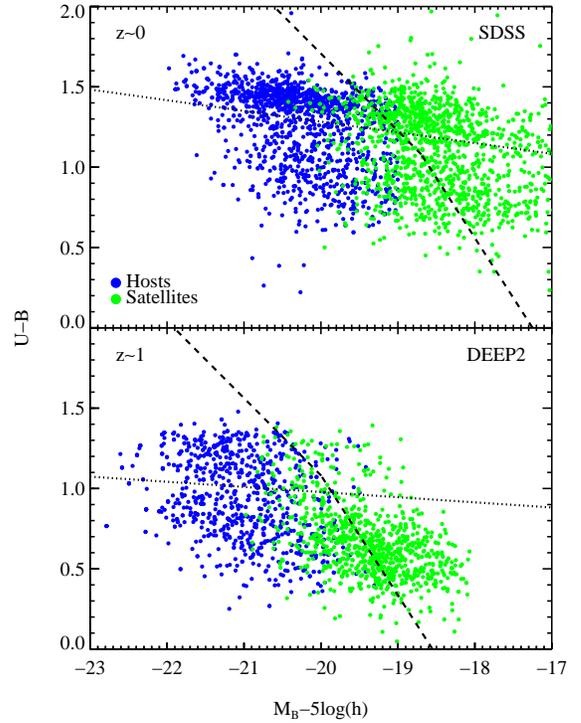}
\vspace{0.5cm}
\caption{Color-magnitude diagram for host galaxies (blue/black points)
  and satellites (green/grey points).  Galaxies drawn from the SDSS at
  $z\sim0$ are plotted in the top panel, while the bottom panel shows
  galaxies at $z\sim1$ from the DEEP2 survey.  The top panel contains
  only $25$\% of the total number of objects for clarity.  The dashed
  broken line defines the completeness limit at $z=1$ (bottom panel),
  and a similar limit at $z=0$ (top panel), where in this case the
  line has been shifted to the right according to the estimated
  evolution in $M_B^\ast$ from $z=1$ to $z=0$.  The dotted line in
  each panel indicates our division between red and blue galaxies.}
\vspace{0.5cm}
\label{f:cmr}
\end{figure}

The DEEP2 Galaxy Redshift Survey \citep{Davis03} has gathered optical
spectra for $\sim40,000$ galaxies in the redshift range $0.7<z<1.4$
using the DEIMOS spectrograph \citep{Faber03} on the Keck II 10-m
telescope.  The survey spans a comoving volume of $\sim5\times10^6$
$h^{-3}$ Mpc$^3$, covering $\sim3$ deg$^2$ over four widely separated
fields.  Target galaxies were selected using $BRI$ imaging from the
CFHT telescope down to a limiting magnitude of $R=24.1$
\citep{Coil04b}.  In three of the four fields we also use apparent
colors to exclude objects likely to have $z<0.7$.  This pre-selection
greatly enhances our efficiency for targeting galaxies at high
redshift \citep{Faber06}.  Due to the high spectral resolution
($R\sim5,000$) and excellent sky subtraction provided by the DEIMOS
spectrograph and DEEP2/DEIMOS data reduction pipeline (Cooper et al.,
in prep), our rms redshift errors are $\sim30$ km s$^{-1}$ determined
from repeated observations.  Details of the DEEP2 observations,
catalog construction, and data reduction can be found in
\citet{Davis03}, \citet{Coil04b}, and \citet{Davis05}.  Restframe
$U-B$ colors and absolute $B$-band magnitudes have been derived as
described in \citet{Willmer06}.  The parent DEEP2 catalog includes
$21,184$ galaxies with high-quality redshifts in the redshift interval
$0.70<z<1.2$.

Stellar masses for DEEP2 galaxies are estimated in the following way.
For the subset of DEEP2 galaxies for which there exists $K_s$-band
imaging, \citet{Bundy06} has determined stellar masses based on the
methodology outlined in \citet{Kauffmann03a} with a Chabrier IMF.  We
then use an empirically derived relation between restframe $UBV$
colors and stellar mass (C.N.A. Willmer, private communication) to
obtain stellar masses for DEEP2 galaxies that do not have $K_s$-band
imaging.  The stellar masses obtained in this way agree well with the
stellar mass estimates obtained for DEEP2 galaxies from the
\texttt{kcorrect v4.1.4} routine (cf. the previous section), as
expected since both methods use the same IMF.  Since the same IMF is
used, we do not expect any systematic offset between the stellar mass
estimates at $z\sim0$ and $z\sim1$.

As with the SDSS data, DEEP2 galaxies are split into red and blue
populations based on the valley in the color-magnitude diagram.  These
two populations are divided in a manner identical to
\citet{Willmer06} using the following color-cut:
\begin{equation}
U-B = -0.032 \,(M_B+21.63) +1.03.
\end{equation}

The DEEP2 survey spectroscopically targets $\sim$60\% of objects that
pass the apparent magnitude and color cuts mentioned above.  Of those
targeted objects, we are able to secure redshifts for $>70$\%.
Follow-up observations have shown that $\sim15$\% of the targets are
objects at $z>1.5$ and fail to yield redshifts from DEEP2 for that
reason (C. Steidel, private communication).  We therefore have
successful redshifts for $60$\%$\times85$\%$=51$\% of all galaxies at
$0.7<z<1.4$ in the surveyed fields with apparent magnitude of
$R<24.1$.

\begin{figure}
\centering
\plotone{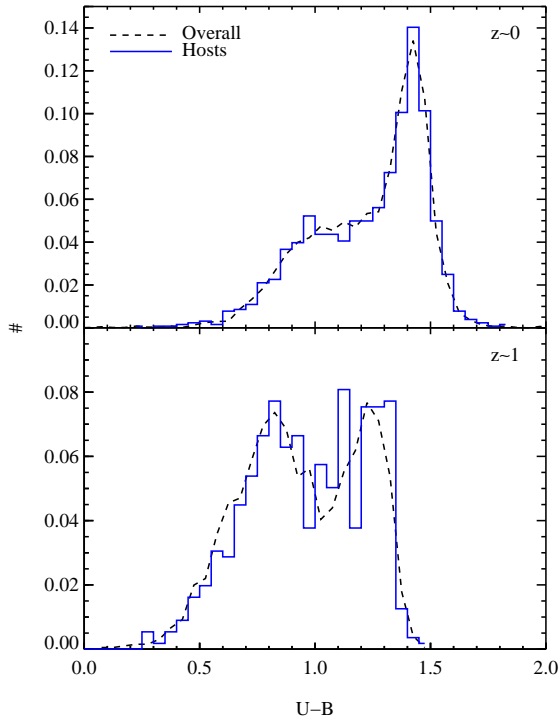}
\vspace{0.5cm}
\caption{$U-B$ colors for host galaxies (\emph{solid lines}) compared
  to all galaxies of similar luminosities and redshifts (\emph{dashed
    lines}), for galaxies at $z\sim0$ (\emph{top panel}) and $z\sim1$
  (\emph{bottom panel}).  The comparison sample was restricted to have
  the same redshift and luminosity distributions as the host galaxy
  sample. The histograms are plotted in units such that the integral
  under each curve is one.}
  \vspace{0.5cm}
\label{f:host_colors}
\end{figure}

\subsection{Defining Uniform Samples from SDSS and
  DEEP2}\label{s:compare}

Every galaxy survey is unique; unfortunately, that makes comparison
between surveys difficult.  In our case, there are two separate
effects that must be taken into account in order to compare the DEEP2
and SDSS galaxy surveys fairly.  The first issue is the differing
sampling rates; this is trivial to resolve, however.  As mentioned in
previous sections, the completeness of the DEEP2 and SDSS surveys are
$\sim50$\% and $\sim90$\%, respectively.  This difference, if not
accounted for, would result in a much larger fraction of falsely
isolated host galaxies in the DEEP2 sample (i.e., galaxies that appear
isolated in the spectroscopic catalog but are not truly isolated) and
could bias the resulting satellite velocity dispersion profile with
respect to the SDSS sample.

In order to mitigate this difference we randomly dilute the SDSS
redshift catalog to the same completeness as DEEP2 (that is, $40$\% of
SDSS galaxies are randomly removed from the catalog).  Extensive tests
with mock catalogs have confirmed that for the DEEP2 survey the
incompleteness is very close to uniform as a function of
$3$-dimensional galaxy over-density \citep{Gerke05,Cooper05a}, hence
this simple random dilution is sufficient for our purposes.  As shown
in Appendix \ref{s:contam}, this dilution results in a $\sim30$\%
\emph{increase} in the estimated virial masses of host galaxies
compared to the complete (undiluted) SDSS.  This should be kept in
mind when considering the results in $\S$\ref{s:results}.

The second effect primarily impacts the satellite population.  While
both surveys select targets that are brighter than an \emph{apparent}
$R$-band magnitude limit, in the DEEP2 survey this selection
corresponds to an approximately \emph{restframe} $B$-band selection at
$z\sim0.7$ and approximately \emph{restframe} $U$-band at $z\sim1$,
while in the SDSS this selection corresponds closely to restframe
$R$-band.  This means that, for the DEEP2 survey, as one considers
fainter objects, redder galaxies will drop out of the survey at a
brighter $M_B$ than bluer galaxies.  This selection effect is well
understood \citep{Willmer06}, and is accounted for it in the following
way.

Figure \ref{f:cmr} shows the color-magnitude distribution for all
host galaxies (blue/black points) and their associated satellites (red
points).  Results for host and satellite galaxies at $z\sim0$ are
shown in the top panel, while results for $z\sim1$ are in the bottom
panel.  The dotted lines indicate our division into red and blue
populations.  Note the increase in the number of red host galaxies
between $z\sim1$ and $z\sim0$ at bright luminosities, where our data
are complete.  The increasing abundance of red galaxies with time has
been studied in detail elsewhere \citep{Bell04,Faber05,Willmer06}.

At a specified redshift, the apparent DEEP2 $R$-band magnitude limit
can be modeled by a broken line in the color-magnitude diagram
(cf. Figure \ref{f:cmr} here and Figure 4 in \citet{Willmer06}; see
also \citet{Gerke06a} for a more detailed discussion of this effect).
In particular, the broken line in Figure \ref{f:cmr} can be used to
define a volume limited sample at $z\leq1$ that uniformly samples
color-magnitude space; when selecting such samples we use only DEEP2
galaxies with $z\leq1$ and include only galaxies that are brighter
than this line in color-magnitude space.  Note that the broken line is
a function of redshift.  Moreover, the dearth of faint red galaxies in
the bottom panel is attributable to this $R$-band selection effect.
For the sample at $z\sim0$ the broken line is shifted according to the
observed evolution in the $B$-band luminosity function, $M_B^\ast(z)=
M_B^\ast(z=0) - 1.37z$, which is approximately independent of color
\citep{Faber05}.  Thus, selecting galaxies brighter than this broken
line ensures that SDSS and DEEP2 are similarly complete \emph{relative
  to $M_B^\ast$} as a function of both luminosity and color at all
redshifts $z\leq1$.

For host galaxies this selection effect is not particularly important
because host galaxies are all rather luminous (by definition, since
these galaxies must be bright enough to have satellite galaxies that
are at least $\delta M_B$ magnitudes fainter).  For such bright galaxies
both very red and very blue host galaxies are included in
both surveys over the full redshift ranges we consider.  This is
apparent in Figure \ref{f:cmr}, where nearly all of the blue/black points in
both panels (representing host galaxies) are to the left of the broken
dashed line.

Unfortunately, accounting for the apparent $R$-band selection effect
drastically reduces the number of available satellite galaxies, as can
been seen in Figure \ref{f:cmr}, where most of the green/grey points
(representing satellite galaxies) are to the right of the broken
dashed line.  In order to increase our statistics, when measuring
velocity dispersions in $\S$\ref{s:results} we do not account for this
selection effect in the manner mentioned above.  Assuming that the
measured velocity dispersion does not depend on \emph{satellite}
properties, then the velocity dispersion should be insensitive to this
selection effect, since including this effect will only decrease the
total number of host galaxies but will not preferentially select one
type of host galaxy (e.g., bright or red) over another\footnote{The
  situation may not be this simple if host galaxy properties are
  strongly correlated with satellite properties.  For example, if red
  satellites preferentially exist around red host galaxies, then by
  missing red satellites we would be implicitly missing red host
  galaxies.  Such a correlation between central and satellite galaxy
  properties has recently been observed at $z\sim0$
  \citep{Weinmann06}, though the signal is not large for the types of
  systems explored here.  Indeed, the fraction of red host galaxies
  does not significantly change when including or neglecting the
  $R$-band effect.}.  However, when comparing host and satellite
properties between DEEP2 and SDSS in $\S$\ref{s:props}, we account for
this selection effect since it strongly impacts the satellite
population.

\section{Properties of Host Galaxies and their Associated
  Satellites}\label{s:props}
  
In this section we present several salient properties of host and
satellite\footnote{Here we are actually presenting the ensemble
  properties of true satellites \emph{and} interloper galaxies.
  Unless explicitly stated otherwise, we refer to the combination of
  true satellites and interlopers as ``satellites'' hereafter.} galaxies
and investigate the evolution of these properties from $z\sim1$ to
$z\sim0$.  Differences in the selection effects in DEEP2 and SDSS are
taken into account as described in the previous section.

\begin{figure}
\centering
\plotone{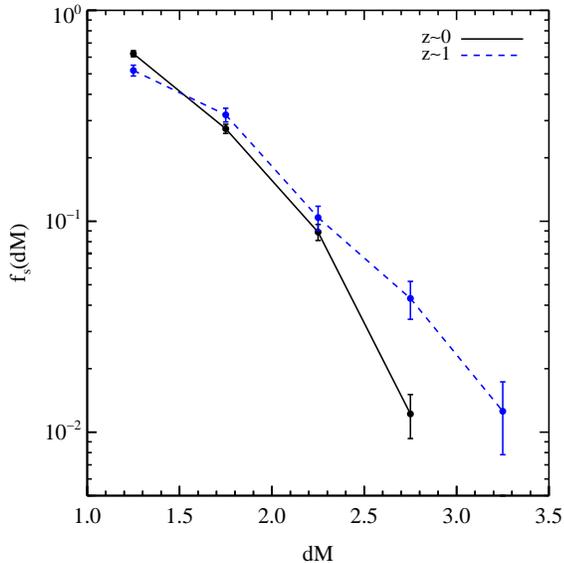}
\vspace{0.5cm}
\caption{Fraction of satellites as a function of $dM\equiv
  M_B^{\rm{sat}}-M_B^{\rm{host}}$ in sample $A$.  Satellite-host systems
  at $z\sim0$ (\emph{solid line}) are compared to systems at $z\sim1$
  (\emph{dashed line}). Selection differences between the parent
  catalogs from which these systems were extracted are carefully
  accounted for using the ``$R$-band cut'' (see $\S$\ref{s:compare}
  for details).}
  \vspace{0.5cm}
\label{f:dm}
\end{figure}

Table \ref{t:params} lists the different search criteria used for
identifying satellites and host galaxies.  Our fiducial sample is $A$;
this section and the next contains results using those search criteria
only.  These search parameters are similar to ones adopted in previous
work \citep{McKay02,Prada03, Conroy05a}.  In Appendix \ref{s:test_mass}, we
demonstrate that our results are robust to the particular choice of
parameters.

Table \ref{t:data} contains information on the host and satellite
samples using search criteria $A$.  The Table includes the redshift
and magnitude ranges over which we search for hosts and satellites,
the median redshift of each sample of host galaxies, the number of
hosts and satellites found in each sample, and the mean satellite
luminosity.  This information is tabulated for samples restricted to
the region of color-magnitude space where the parent catalogs are
complete at both epochs (the samples used in this section, labeled in
the table ``with $R$-band cut''), and samples that include regions of
color-magnitude space where the parent catalogs are not complete (the
samples used in the following section, labeled ``without $R$-band cut''
in the table).

In order to quantify the distribution of $U-B$ colors for host
galaxies, Figure \ref{f:host_colors} plots histograms of $U-B$
color at $z\sim1$ (top panel) and $z\sim0$ (bottom panel).  Each
distribution is normalized such that its integral is unity.  The
sample of host galaxies (solid lines) is compared to an ``overall''
sample of galaxies (dashed lines).  The comparison sample has been
constructed such that it samples the redshift-luminosity plane with
the same density as the host galaxies.  It is evident that host
galaxies have a distribution in $U-B$ colors comparable to all
galaxies at the same luminosity at both $z\sim1$ and $z\sim0$, and
that host galaxies at $z\sim0$ are redder than host galaxies at
$z\sim1$ only insofar as the overall galaxy population is redder at
$z\sim0$ compared to $z\sim1$.  One can also clearly see the growth in
the abundance of red galaxies between $z\sim1$ and $z\sim0$, as
noticed in previous studies \citep{Bell04,Faber05,Willmer06}.

The samples used in Figure \ref{f:host_colors} have the $R$-band cut
taken into account.  However, the full samples (neglecting the
$R$-band cut) display nearly identical host $U-B$ color distributions
when compared to the $R$-band cut samples.  This indicates that the
full samples, which are used in $\S$\ref{s:results}, do not contain
any \emph{artificial} changes in the $U-B$ color distribution of
galaxies with time.

Generating a fair comparison sample is a requisite for interpreting
the distribution of colors of host galaxies.  Since host galaxies are
in general much brighter than an average galaxy (i.e., a galaxy drawn
at random from the full survey), without a fair comparison sample we
would have falsely concluded that host galaxies are redder than the
average galaxy.  These selection effects are exacerbated in the DEEP2
sample, where the red galaxy population is a strong function of
redshift due both to the apparent $R$-band limit of the survey and to
evolution in red number density, but are effectively mitigated with a
proper comparison sample.

Figure \ref{f:dm} plots the distribution, $f_s(dM)$, of magnitude
differences, $dM\equiv M_B^{\rm{sat}}-M_B^{\rm{host}}$ at $z\sim1$
(dashed line) and $z\sim0$ (solid line).  Error bars denote $1\sigma$
Poisson uncertainties.  We conclude from these distributions that the
average host galaxy at $z\sim1$ has fainter satellites than an average
host galaxy at $z\sim0$.  This conclusion is unchanged if we
separately consider $f_s(dM)$ in bins of host galaxy luminosity.  The
average $dM$ for each sample reflects this difference as well, though
less strikingly: at $z\sim1$ $\langle dM \rangle=1.60$ while at
$z\sim0$ $\langle dM \rangle=1.46$.

Note that since these distributions of $f_s(dM)$ are normalized to the
total number of satellites at each redshift, these results are
insensitive to the presence of interlopers unless interlopers have a
$dM$ distribution that varies with redshift.  Although such a scenario
seems particularly nefarious, it cannot explicitly be ruled out.

In addition we have investigated the sensitivity of these results to
our assumed evolution in the luminosity function.  Throughout we have
assumed that the characteristic scale of the luminosity function
evolves as $M_B^\ast(z) = M_B^\ast(z=0) - 1.37z$ independent of color
\citep{Faber05}.  The evolution in the luminosity function is
important here because it determines how the dashed line in Figure
\ref{f:cmr} evolves with redshift, which in turn defines the samples
used in this section.  If the amount of evolution in $M_B^\ast$ is
varied by $\pm0.3z$ \citep[which is the $1\sigma$ uncertainty in the
evolution of $M_B^\ast$ as reported by][]{Faber05}, the qualitative
result of Figure \ref{f:dm}, namely that DEEP2 satellites are on
average fainter with respect to their host luminosities than SDSS
satellites, remains unchanged.  However, the case of maximal
evolution, where $M_B^\ast$ evolves by $1.67$ magnitudes per unit
redshift, results in a significantly less striking difference at large
$dM$, for the obvious reason that this maximal evolution in $M_B^\ast$
allows many more faint satellite galaxies to be included in the sample
at $z\sim0$ (in essence, the dashed line in the top panel of Figure
\ref{f:cmr} moves to the right).  Thus, if the difference in satellite
properties seen in Figure \ref{f:dm} is not real, then evolution in
$M_B^\ast$ is even stronger than found by \citet{Faber05}.

Moreover, dynamical friction acting on the $z\sim1$ host-satellite
population would tend to produce a trend \emph{opposite} to what is
observed here.  Because dynamical friction is most efficient when it
acts between objects of comparable mass, it will cause the brightest
satellites to sink toward the host preferentially, resulting in
comparably \emph{more} fainter satellites at $z\sim0$ than at
$z\sim1$, contrary to our observations.

\begin{figure}
\centering
\plotone{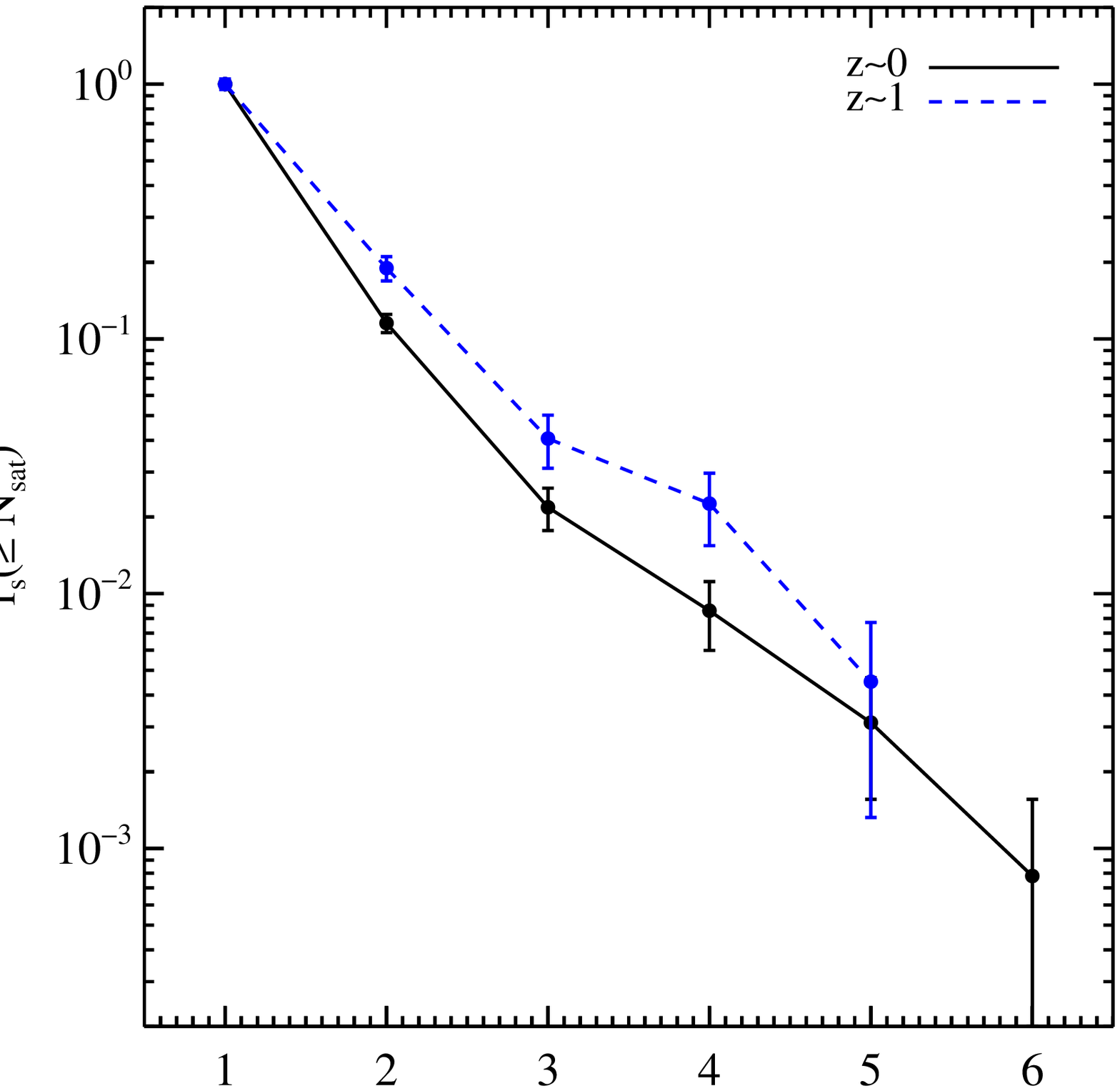}
\vspace{0.5cm}
\caption{Fraction of host galaxies that have $\geq N_{\rm{sat}}$
  satellites, as a function of $N_{\rm{sat}}$.  As in the previous
  figure, we use search criteria $A$ and account for differences
  between the parent catalogs at $z\sim1$ and $z\sim0$ using the
  ``$R$-band cut'' as described in $\S$\ref{s:compare}.}
\label{f:nsat}
\vspace{0.5cm}
\end{figure}

In Figure \ref{f:nsat} we plot the fraction of host galaxies with at
least $N_{\rm{sat}}$ satellites, $f_s(\geq N_{\rm{sat}})$, as a
function of $N_{\rm{sat}}$, both at $z\sim1$ (dashed line) and
$z\sim0$ (solid line).  Errors reflect Poisson uncertainties.  It is
evident that, when comparing volume limited samples at $z\sim1$ and
$z\sim0$ which are similarly complete in color-magnitude space, host
galaxies at high redshift are associated with more satellites than
host galaxies at low redshift (although the difference is weak, see
below).  This is illustrated alternatively by considering the average
number of satellites per host galaxy: $\langle N \rangle=1.26$ at
$z\sim1$ and $\langle N \rangle=1.15$ at $z\sim0$.

In this case interlopers have a more direct impact. In
$\S$\ref{s:results} we show that the interloper fraction decreases
from $\sim21\pm6$\% at $z\sim1$ to $\sim16\pm4$\% at $z\sim0$ (so that
within $1\sigma$ the interloper fraction is constant with redshift).
Therefore, the changes in the average number of satellites per galaxy
could be due to changes in interloper contamination with $z$ at
$<1\sigma$.  Uncertainties in the evolution of $M_B^\ast$ from
$z\sim1$ to $z\sim0$ also strongly impact the results shown in Figure
\ref{f:nsat} because a larger/smaller evolution in $M_B$ than what is
assumed here will result in more/less faint galaxies in the $z\sim0$
``$R$-band cut'' sample, which will in turn result in more/less
satellites at $z\sim0$ compared to $z\sim1$.  In short, these results
are unfortunately too sensitive to various assumptions and
uncertainties to provide robust conclusions regarding the differential
evolution in satellite numbers between $z\sim1$ and $z\sim0$.

\begin{deluxetable}{lll}
\tablecaption{Summary of Samples} \tablehead{
  \colhead{} &
  \colhead{SDSS} & 
  \colhead{DEEP2} }
\startdata
redshift range  & $0.01<z<0.10$ & $0.7<z<1.2$ \\
$\langle z\rangle$ & 0.06 & 0.84 \\
Host $M_B-5\rm{log}$$(h)$ range  & $<-19$ & $<-20$ \\
\\
\emph{without R-band cut:} & & \\
Total Sample Size & $102,656$ & $21,184$ \\
Satellite $\langle M_B-5\rm{log}$$(h)\rangle$ & $-18.3$ & $-19.4$ \\
$N_{\rm{sat}}$ & 5414 & 1007 \\
$N_{\rm{host}}$ & 3431 & 755 \\
\\
\emph{with R-band cut:} & & \\
Total Sample Size & $56,397$ & $12,887$ \\
Satellite $\langle M_B-5\rm{log}$$(h)\rangle$ & $-19.0$ & $-19.6$ \\
$N_{\rm{sat}}$ & 1475 & 554 \\
$N_{\rm{host}}$ & 1283 & 440 \\
\enddata
\tablecomments{The host and satellite galaxies contained in these
  samples were obtained using search criteria $A$ (cf. Table
  \ref{t:params}).  The total sample size for the SDSS survey refers
  to the survey after it has been diluted by $40$\% to match the
  completeness of DEEP2. \vspace{0.5cm}}
\label{t:data}
\end{deluxetable}

\section{Velocity Dispersion Measurements \& Derived Virial
  Masses}\label{s:results}

We now present the measured satellite velocity dispersion, $\sigma$,
as a function of host galaxy redshift, luminosity, stellar mass, and
color.  In addition, we derive virial mass-to-light ratios
($M_{200}/L_B$) and virial-to-stellar mass ratios ($M_{200}/M_\ast$)
at $z\sim1$ and $z\sim0$.  In this section the $R$-band selection
effect is \emph{not} accounted for when comparing samples at $z\sim1$
and $z\sim0$; see $\S$\ref{s:compare} for details.  Results are for
host and satellite galaxies identified according to search criteria
$A$ (cf. Table \ref{t:params}).  Since both SDSS and DEEP2 are
constructed to have the same completeness ($\sim50$\%), all virial
masses quoted herein are $\sim30$\% higher than they would be had
$100$\% complete surveys been used.  This is attributable to systems
that only appear isolated in the diluted sample, but in fact reside
within more massive halos with multiple bright companions
(see Appendix \ref{s:contam}).

Appendix \ref{s:test_mass} demonstrates that the results presented in
this section are robust to various assumptions, including the specific
search criteria used to define the samples and the velocity anisotropy
and concentration of the satellites orbiting within their host
galaxy's dark matter halo.  Appendix \ref{s:appraisal} presents an
assessment of and motivation for the broader methodological
assumptions inherent in using the motions of satellite galaxies to
extract host galaxy halo masses.

\begin{figure*}
\centering
\plottwo{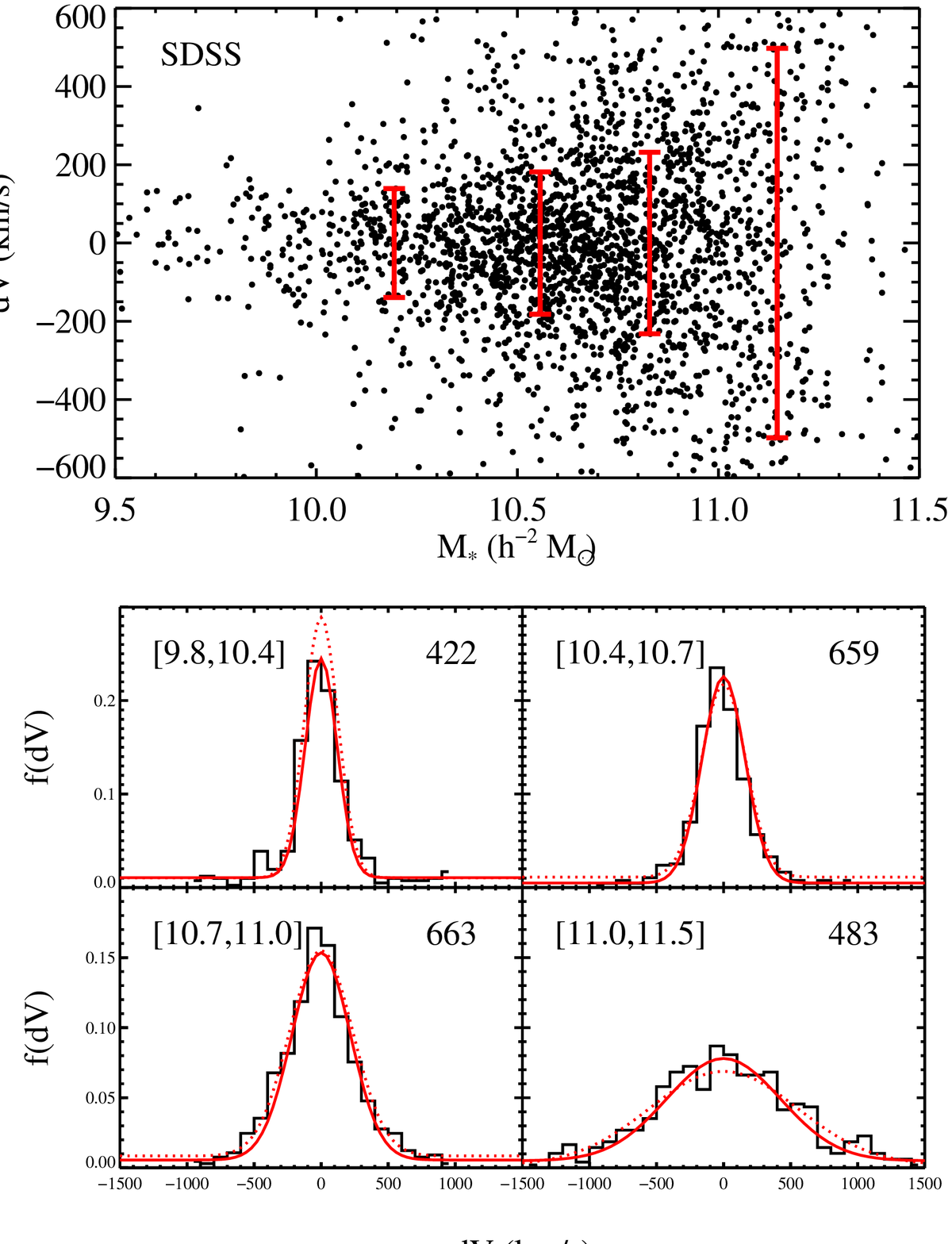}{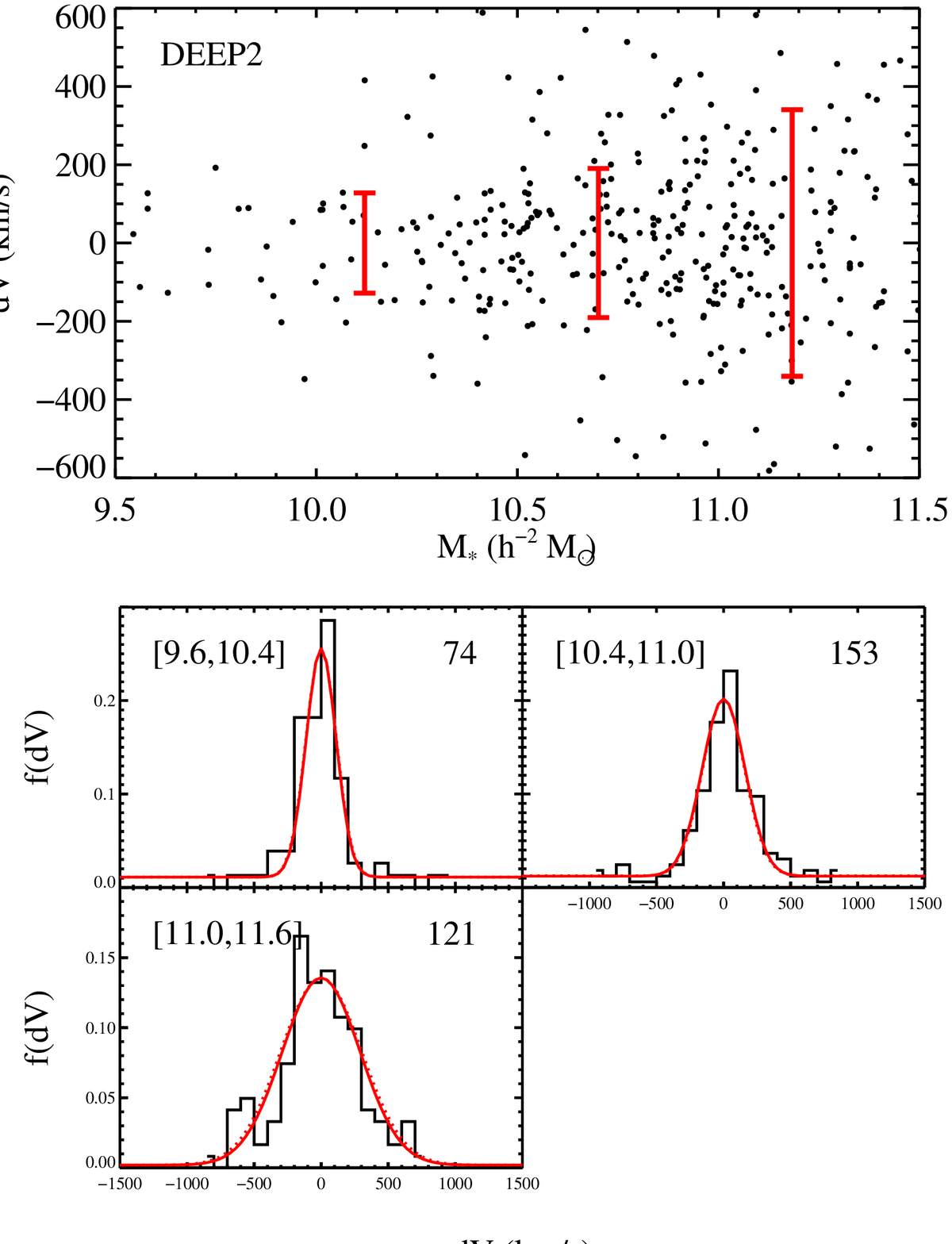}
\vspace{0.5cm}
\caption{\emph{Left panels:} The top panel plots the relative
  line-of-sight satellite velocity, $dV$, as a function of host
  stellar mass, $M_\ast$, for all satellites within $R_p=[20,150]$
  $h^{-1}$ kpc at $z\sim0$ from sample $A$.  The solid vertical lines
  indicate the full width at half maximum taken from the measured
  dispersion within four bins in host stellar mass.  The bottom panels
  display the $dV$ distribution in bins of stellar mass normalized to
  the total number of satellites within the bin (stellar mass bin
  width is displayed in the top left corner in units of
  $\rm{log}(M_\ast$ $h^{-2} M_{\sun}$)).  The top right corner notes
  the number of satellites within each bin.  We over-plot the best fit
  Gaussian+constant model weighted by host galaxy (used in the
  following analysis; \emph{solid lines}) and weighted by satellite
  galaxy (\emph{dotted lines}).  \emph{Right panels:} Same as the left
  panels, now for data at $z\sim1$.}
  \vspace{0.5cm}
\label{f:ml_raw}
\end{figure*}
 
The nominal redshift limit of the SDSS parent catalog has been
extended from $z=0.1$ to $z=0.2$ for the highest stellar mass bin,
doubling the number of satellites in that bin.  Separately measuring
the dispersion in this stellar mass bin for the fiducial sample with
$z\leq0.1$ and the extended sample with $z\leq0.2$ results in
differences within $10$ km s$^{-1}$.  In other words, adding these
higher redshift data does not appear to bias the resulting
measurement, but it significantly decreases the errors on the
dispersion due to the increased number of satellites.  Increasing the
redshift limit of the parent catalog for the other samples has little
effect since the apparent magnitude limit of the SDSS yields few faint
satellite galaxies at higher redshifts.

\subsection{Results as a Function of Host Galaxy Luminosity and
  Stellar Mass}

\begin{deluxetable}{llllll}
\tablecaption{Velocity dispersions and halo masses 
  of host galaxies as a function of $M_B$} 
\tablehead{
  \colhead{$M_B$ bin} &
  \colhead{$\langle M_B\rangle$} &
  \colhead{color} &
  \colhead{$N_{\rm{sats}}$} &
  \colhead{$\sigma$} & 
  \colhead{$M_{200}/L_B$} \\
 \colhead{} &
  \colhead{} & 
  \colhead{} &
  \colhead{} &
  \colhead{km s$^{-1}$} & 
  \colhead{$h$ $M_\Sun/L_\Sun$} }
\startdata

  \multicolumn{6}{c}{\hbox to 90pt{\hfill  \emph{Results at $z\sim0$:}
\hfill}}  \\
$[-19.25,-19.75]$ & $-19.5$ & all & $332$ & $132^{+11}_{-11}$ & $228^{+92}_{-61}$ \\
$[-19.75,-20.25]$ & $-20.0$ & all & $505$ & $144^{+11}_{-9}$ & $180^{+73}_{-36}$ \\
$[-20.25,-20.75]$ &  $-20.5$ & all & $657$ & $188^{+13}_{-11}$ & $254^{+58}_{-66}$ \\
$[-20.75,-21.50]$ &  $-19.5$ & all & $413$ & $235^{+26}_{-24}$ & $277^{+118}_{-110}$ \\

$[-19.25,-20.00]$ &  $-19.7$ & red & $299$ & $159^{+20}_{-18}$ & $357^{+160}_{-150}$ \\
$[-20.00,-20.50]$ &  $-20.3$ & red & $366$ & $215^{+15}_{-15}$ & $454^{+108}_{-141}$ \\
$[-20.50,-21.50]$ &  $-20.9$ & red & $475$ & $258^{+24}_{-22}$ & $384^{+143}_{-124}$ \\

$[-19.25,-20.00]$ & $-19.7$ & blue & $267$ & $118^{+11}_{-11}$ & $166^{+56}_{-57}$ \\
$[-20.00,-20.50]$ &  $-20.3$ & blue & $247$ & $139^{+13}_{-11}$ & $143^{+49}_{-44}$ \\
$[-20.50,-21.50]$ &  $-20.8$ & blue & $215$ & $186^{+18}_{-18}$ & $188^{+66}_{-68}$ \\

\\
 \multicolumn{6}{c}{\hbox to 90pt{\hfill  \emph{Results at $z\sim1$:}
\hfill}}  \\
$[-19.50,-20.75]$ & $-20.4$ & all & $79$ & $118^{+20}_{-18}$ & $71^{+45}_{-37}$ \\
$[-20.75,-21.50]$  & $-19.5$ & all & $154$ & $153^{+22}_{-20}$ & $63^{+34}_{-29}$ \\
$[-21.50,-23.00]$  & $-21.9$ & all & $117$ & $272^{+29}_{-40}$ & $130^{+84}_{-51}$ \\

$[-20.50,-22.00]$ & $-21.3$ & red & $133$ & $231^{+46}_{-37}$ & $151^{+145}_{-79}$ \\

$[-20.50,-22.00]$ &  $-21.2$ & blue & $149$ & $144^{+20}_{-18}$ & $49^{+27}_{-21}$ \\
\enddata
\tablecomments{Absolute $B$-band magnitudes, $M_B$, are quoted as
  $M_B-5\rm{log}$$(h)$.  Recall that masses and dispersions are
  \emph{systematically overestimated} (by $\sim30$\% in mass) due to
  incompleteness effects (cf. Appendix \ref{s:contam} and
    $\S$\ref{s:compare}). \vspace{0.5cm}}
\label{t:res1}
\end{deluxetable} 

In Figure \ref{f:ml_raw} we present the relative velocity of
satellites, $dV$, as a function of host galaxy stellar mass, $M_\ast$,
for all satellites within $R_p=[20,150]$ $h^{-1}$ kpc at $z\sim0$ (top
left panel) and $z\sim1$ (top right panel)\footnote{As we show in
  Appendix \ref{s:test_mass}, the results presented in this section do
  not change significantly when using larger, smaller, or more bins in
  $R_p$.  Our fiducial bin size is motivated by the fact that
  interloper fractions increase with increasing $R_p$, so although one
  includes more true satellites with a larger maximum $R_p$, the
  larger interloper fraction results in no significant improvement in
  the dispersion measurement.}.  Note again that the samples used in
this and the following figures do not account for the different survey
selection effects between DEEP2 and SDSS (which we argued in
$\S$\ref{s:compare} should not impact the results in this section) and
is hence a super-set of the samples used in $\S$\ref{s:props}.  One
can see clearly that the satellite velocity dispersion is increasing
with increasing host stellar mass.  The solid vertical lines indicate
the full width at half maximum given by our dispersion measurements,
in bins of host stellar mass.  The bottom cluster of panels shows the
distribution of relative satellite velocities in bins of stellar mass.
The smooth lines indicate our best fit Gaussian+constant models for
host weighting (solid lines) and satellite weighting (dotted lines).
The difference between weighting schemes becomes increasingly
important for higher stellar mass bins, since the hosts in these bins
have on average more satellites.  We only display results as a
function of stellar mass for simplicity, but we also fit for $\sigma$
in bins of $M_B$.

\begin{deluxetable}{llllll}
\tablecaption{Velocity dispersions and halo masses of host galaxies
  as a function of $M_\ast$} 
\tablehead{
  \colhead{$M_\ast$ bin} &
  \colhead{$\langle M_\ast\rangle$} &
  \colhead{$N_{\rm{sats}}$} &
  \colhead{$\sigma$} & 
  \colhead{$M_{200}/M_\ast$ $h$} \\
 \colhead{} &
  \colhead{} & 
  \colhead{} &
  \colhead{km s$^{-1}$} & 
  \colhead{} }
\startdata

 \multicolumn{5}{c}{\hbox to 90pt{\hfill  \emph{Results at $z\sim0$:}
\hfill}}  \\
$[9.8,10.4]$  & $10.2$   & $422$ & $118^{+9}_{-9}$ & $105^{+27}_{-31}$ \\
$[10.4,10.7]$ & $10.6$  & $659$ & $155^{+11}_{-9}$ & $83^{+32}_{-15}$ \\
$[10.7,11.0]$  & $10.8$  & $663$ & $197^{+15}_{-15}$ & $81^{+36}_{-19}$ \\
$[11.0,11.5]$ & $11.1$  & $483$ & $423^{+59}_{-48}$ & $333^{+146}_{-155}$ \\

\\
 \multicolumn{5}{c}{\hbox to 90pt{\hfill  \emph{Results at $z\sim1$:}
\hfill}}  \\
$[9.6,10.4]$ &  $10.1$  & $74$ & $109^{+22}_{-15}$ & $69^{+61}_{-35}$ \\
$[10.4,11.0]$ & $10.7$ & $153$ & $162^{+24}_{-20}$ & $57^{+29}_{-28}$ \\
$[11.0,11.6]$ & $11.2$ & $121$ & $290^{+31}_{-40}$ & $77^{+46}_{-30}$ \\

\enddata
\tablecomments{All stellar masses, $M_\ast$, are quoted in units of
  $\rm{log}(M_\ast$ $h^{-2} M_{\sun}$).  Recall that masses and
  dispersions are \emph{systematically overestimated} (by $\sim30$\%
  in mass) due to incompleteness effects (cf. Appendix
  \ref{s:contam} and $\S$\ref{s:compare}). \vspace{0.5cm}}
\label{t:res2}
\end{deluxetable}

\begin{figure*}
\centering
\plottwo{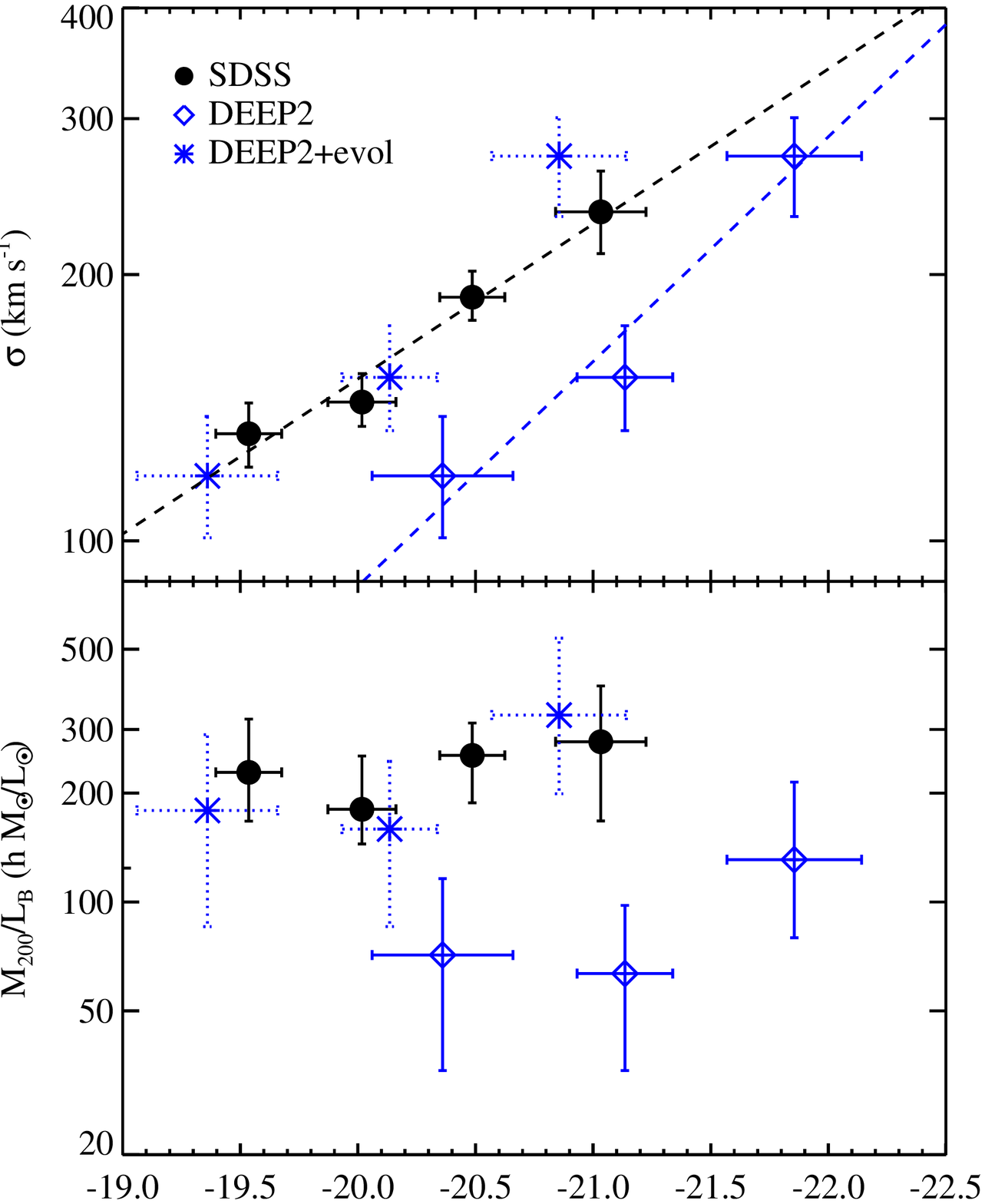}{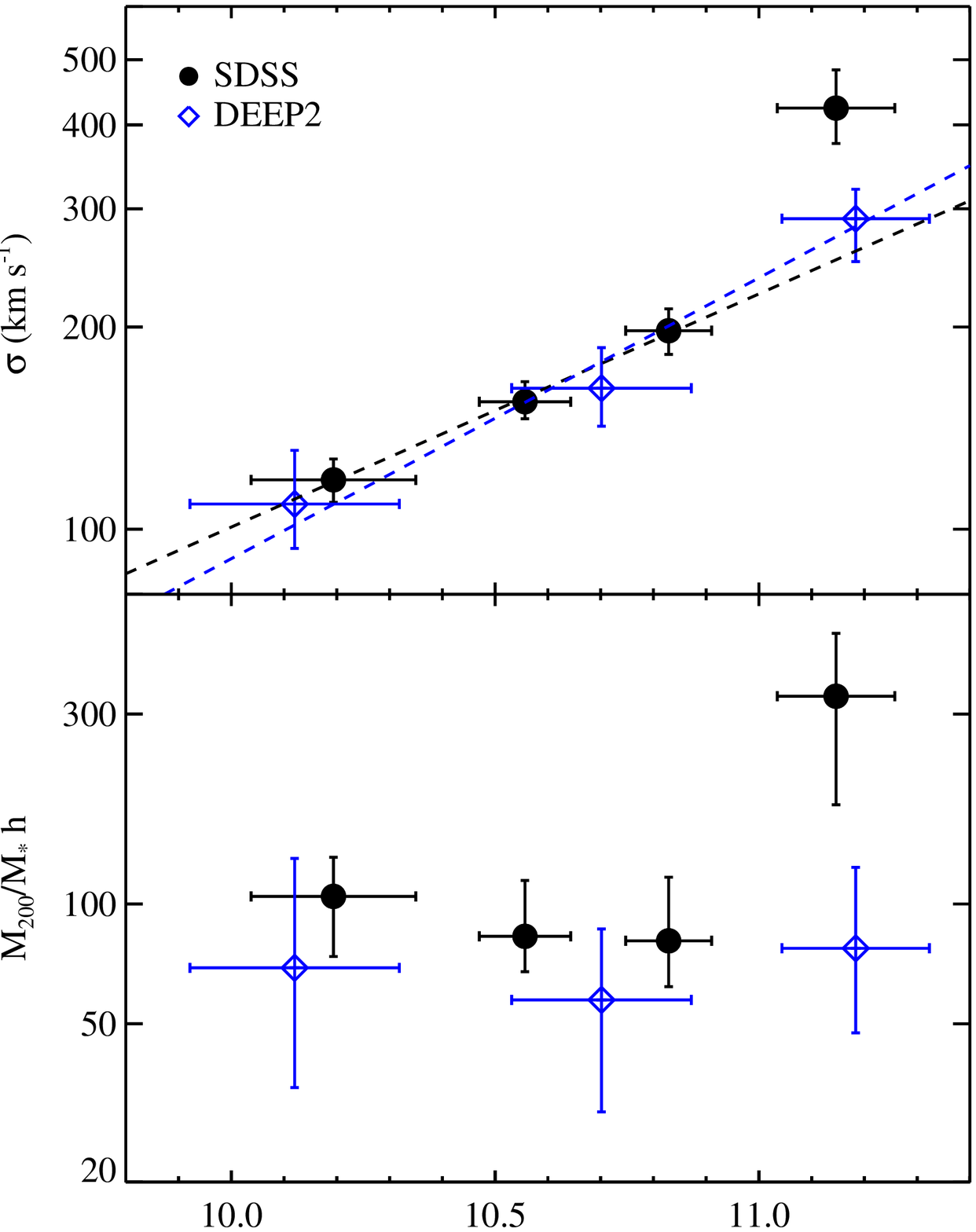}
\vspace{0.5cm}
\caption{\emph{Top panels:} Satellite velocity dispersion, measured
  within $R_p=[20,150]$ $h^{-1}$ kpc, for host galaxies at $z\sim1$
  (\emph{diamonds}) and $z\sim0$ (\emph{circles}) from sample $A$.  In
  the left panel we plot the dispersion as a function of the $B$-band
  absolute magnitude of the host galaxy, while the right panel is as a
  function of stellar mass ($M_\ast$).  In addition we show the
  $z\sim1$ measurements modified by a one magnitude dimming in the
  $B$-band luminosity of host galaxies (\emph{stars}).  The dashed
  lines are from Equations \ref{e:sl} and \ref{e:ml}.  \emph{Bottom
    left panel:} Mass-to-light ratios as a function of host galaxy
  $M_B$.  \emph{Bottom right panel}: Virial-to-stellar mass ratio as a
  function of host galaxy stellar mass.  Since virial mass is defined
  with respect to the redshift-dependent critical density, an increase
  in $M_{200}/M_\ast$ of $1.3$ is expected for an intrinsically
  non-evolving $M_{200}/M_\ast-M_\ast$ relation (the same is true as a
  function of $M_B$).  Symbol types are the same as the top panels.}
\vspace{0.5cm}
\label{f:ml_data}
\end{figure*}

Figure \ref{f:ml_data} contains our main results.  The top panels
present the satellite velocity dispersion measured within a
projected separation of $R_p=[20,150]$ $h^{-1}$ kpc as a function of
$B$-band magnitude (top left panel) and stellar mass (top right panel)
for galaxies at $z\sim1$ (blue diamonds) and $z\sim0$ (black circles).
Thee $1\sigma$ errors on the dispersion measurement were
obtained from the likelihood contours described in
$\S$\ref{s:mass_est} and error bars in the $B$-band magnitude on the
$x$--axis represent the $68$\% range within each magnitude bin.
The information in these figures is also listed in Tables \ref{t:res1}
and \ref{t:res2} where we list, for each bin in host $M_B$ and
$M_\ast$, the number of satellites, mean $M_B$ and
$M_\ast$ of host galaxies, recovered satellite velocity dispersions,
and virial-to-stellar mass and mass-to-light ratios.

In the bottom panels we present the same information, where now the
velocity dispersion measurements have been converted into virial
masses (cf. $\S$\ref{s:mass_est}) and quote our results in terms
of the virial mass-to-light ratio ($M_{200}/L_B$, bottom left panel)
and the virial-to-stellar mass ratio ($M_{200}/M_\ast$, bottom right
panel).  Note that in order to convert the velocity dispersion into a
virial mass we use only the velocity dispersion within $R_p=[20,150]$
$h^{-1}$ kpc rather than attempting to measure the dispersion in
multiple bins; this is due to our limited statistics, especially at
$z\sim1$.

Errors on the virial mass-to-light ratios reflect the errors on the
virial mass, which were obtained from $\chi^2$ minimization between
the observed velocity dispersion and the theoretical dispersion
profile.  Any errors in the luminosity are not included when
calculating the error on $M_{200}/L_B$; we simply use the mean
luminosity within each bin (the median $B$-band magnitude agrees with
the mean to within $\pm0.01$).  Error bars on the $B$-band magnitudes
on the $x$--axis again represent the $68$\% range in each magnitude
bin.  The virial-to-stellar mass ratios are plotted similarly.

We mention in passing that the mean redshift of host galaxies does not
vary strongly across the $B$-band magnitude and stellar mass bins.  At
$z\sim0$ the mean redshift varies from $0.044$ in the faintest bin to
$0.076$ in the brightest bin, while at $z\sim1$ it increases from
$0.80$ to $0.91$ between the faintest to brightest magnitude bins
explored here.

In addition to velocity dispersion measurements, our
maximum-likelihood technique provides an estimate of the interloper
fraction (the number of interlopers divided by the total number of
satellites) for each sample.  The interloper fraction for the $z\sim0$
samples is $\sim16\pm4$\% and is constant (within $1\sigma$) across
the host luminosity and stellar mass bins.  This interloper fraction
at $z\sim0$ is in agreement with previous work.  In particular,
\citet{Prada03} found interloper fractions of $17-20$\% depending on
their sample definition.  At $z\sim1$ the interloper fractions are
noisier due to the smaller number of satellites.  We measure an
average interloper fraction at $z\sim1$ of $\sim21\pm6$\%, again with
little variation across the luminosity and stellar mass bins.  While
the interloper fractions at both $z\sim0$ and $z\sim1$ are consistent
within $1\sigma$, $z\sim1$ host galaxies do have a slightly higher
fraction of interlopers than $z\sim0$ hosts.  A higher interloper rate
at higher redshift might be attributable to the fact that more systems
were still in the process of assembling then, and hence the higher
fraction of interlopers could be reflecting a higher fraction of
systems that have not yet settled into dynamical equilibrium.  Such a
trend should not bias our velocity dispersion and mass estimates
because the interlopers are effectively accounted for in the
dispersion measurements, regardless of their fraction.\footnote{These
  interloper fractions are likely underestimated because we have
  assumed that interlopers are distributed uniformly in $dV$ while
  tests based on mock galaxy catalogs suggest that the interloper
  distribution is more complex \citep{VDB04b,Chen06}.  Based on tests
  with mock catalogs, \citet{VDB04b} found that the assumption of a
  constant distribution of interlopers underestimates the interloper
  fraction by as much as $50$\%.  Note, however, that these details do
  not impact the recovered satellite velocity dispersion and will
  affect the interloper fractions at $z\sim1$ and $z\sim0$ in the same
  way.}

Figure \ref{f:ml_data} includes power-law fits to the observed
$\sigma-L_B$ and $\sigma-M_\ast$ relations:
\begin{equation}
\label{e:sl}
\sigma \propto\left\{ \begin{array}{ll} L_B^{0.4}, & z\sim0 \\
[1mm] L_B^{0.6}, & z\sim1,
\end{array}
\right.
\end{equation}
\begin{equation}
\label{e:ml}
\sigma \propto\left\{ \begin{array}{ll} M_\ast^{0.4}, & z\sim0 \\
[1mm] M_\ast^{0.4}, & z\sim1,
\end{array}
\right.
\end{equation}
with $1\sigma$ errors on the exponents of $0.1$.  We use only the
lower three $\sigma(M_\ast)$ points for the fit at $z\sim0$ because
the dispersion in the highest stellar mass bin seems to deviate
strongly from the relation indicated by the other points.  Although
the $\sigma-L_B$ relation at $z\sim1$ is formally steeper than the
corresponding relation at $z\sim0$, we emphasize that the two slopes
are consistent with one another, and stress that the inferred slope of
the $\sigma-L_B$ relation depends on the range of luminosities probed
and hence the exponents in this relation should in general be treated
with caution (the same can be said of the $\sigma-M_\ast$ relation).

For an NFW density profile with no anisotropy, virial mass is related
to velocity dispersion as $M_{200} \propto \sigma^{2.5}$ at $R_p=100$
$h^{-1}$ kpc (which is roughly the mean satellite $R_p$ used in our
analysis), and thus $M_{200}/L_B \propto \sigma^{2.5}/L_B$.  Hence, at
$z\sim0$, Equation \ref{e:sl} implies that $M_{200}/L_B \propto
L^{2.5\times0.4-1} \sim$ constant, while at $z\sim1$ $M_{200}/L_B
\propto L^{0.5\pm0.3}$.  Notice that in general, if $\sigma\propto
L^\alpha$, $M/L$ will increase with $L$ only when $\alpha> 0.4$.
Similar equations can be derived as a function of host galaxy stellar
mass.  These inferred $M_{200}/L_B$ versus $L_B$ trends are perfectly
consistent with the observed trends in the bottom panels of Figure
\ref{f:ml_data}.  In addition, the slope of our measured $\sigma-L_B$
relation at $z\sim0$ is consistent with previous estimates from
satellite dynamics \citep{McKay02,Prada03, Brainerd05}.

One should keep in mind that the $\sigma-L_B$ and $\sigma-M_\ast$
relations are not directly comparable because a bin in $L_B$ contains
a different fraction of red host galaxies from a similar bin in
stellar mass.  This is due to the fact that stellar mass is much more
strongly correlated with galaxy color than $B$-band absolute
magnitude.  At $z\sim0$ the fraction of host galaxies with red $U-B$
color increases from $48$\% to $77$\% in our lowest to
highest $B$-band magnitude bins, while the red fraction increases from
$26$\% to $91$\% from the lowest to highest stellar mass bins.
Similarly, at $z\sim1$ the fraction of red host galaxies increases
from $33$\% to $55$\% from the faintest to brightest $B$-band
magnitude bin, and from $4$\% to $88$\% for the smallest to largest
stellar mass bins.  The strongly varying fraction of red host galaxies
as a function of host stellar mass makes the comparison and
interpretation of the evolution in $\sigma(M_\ast)$ between $z\sim1$ and
$z\sim0$ simpler because each end of the $\sigma-M_\ast$ relation is
dominated by a single population (red galaxies at the massive end and
blue galaxies at the faint end).

The $\sigma-M_\ast$ and $M_{200}/M_\ast-M_\ast$ relations show little
evolution between $z\sim1$ and $z\sim0$ except for the highest stellar
mass bin.  For host galaxies with $M_\ast\lesssim10^{11}$ $h^{-2}
M_\Sun$ the raw increase in $M_{200}/M_\ast$ is constrained to
$1.4\pm0.9$ (but see below).  For host galaxies with larger stellar
mass the virial-to-stellar mass ratio increases by a factor of $4\pm3$
between these epochs.  These hosts have large dispersions
($\gtrsim300$ km s$^{-1}$) that correspond to massive dark matter
halos ($>10^{13}$ $h^{-1} M_\Sun$), which commonly contain groups of
galaxies, and hence they probably should not be interpreted in the
same way as less massive host galaxies.  We defer a more detailed
discussion of this extreme population to $\S$\ref{s:disc2}.

\begin{figure}
\centering
\plotone{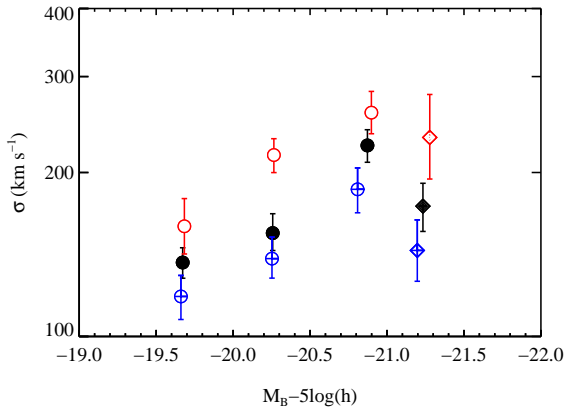}
\vspace{0.5cm}
\caption{Satellite velocity dispersion as a function of host galaxy
  $U-B$ color and $M_B$ for samples at $z\sim0$ (\emph{circles}) and
  $z\sim1$ (\emph{diamonds}).  At fixed $M_B$ there is a clear
  difference between all (\emph{solid, black}), red (\emph{open red}),
  and blue (\emph{open with crosses, blue}) hosts both at $z\sim1$ and
  $z\sim0$.}
\vspace{0.5cm}
\label{f:ml_color}
\end{figure}

When interpreting the evolution in the virial mass-to-light ratio, one
should keep in mind that our \emph{definition} of mass changes with
redshift since the critical density is redshift-dependent.
Specifically, at a fixed velocity dispersion, the virial mass defined
according to a region enclosing a mean density $200$ times the
critical density, $M_{200}$, \emph{increases} by a factor of $\sim1.3$
from $z\sim1$ to $z\sim0$.  A changing virial mass with redshift, even
for a static, intrinsically non-evolving dark matter halo (i.e., a
halo that is not accreting new material), is a generic feature of all
common virial definitions.  In other words, an increase in
$M_{200}/M_\ast$ of $1.3$ between $z\sim1$ and $z\sim0$ is expected
for an intrinsically non-evolving $M_{200}/M_\ast-M_\ast$ relation.

Such an increase is precisely what is observed for host galaxies with
$M_\ast\lesssim10^{11}$ $h^{-2} M_\Sun$, which show a raw increase in
$M_{200}/M_\ast$ of $1.4\pm0.9$.  Hence even this small increase does
not reflect physically interesting changes in host galaxies.  Folding
in the changing virial definition with redshift results in an
``intrinsic'' increase in $M_{200}/M_\ast$ of only $1.1\pm0.7$.  This
differentiation between intrinsic and definitional changes in
$M_{200}/M_\ast$ is supported by the $\sigma-M_\ast$ relation for host
galaxies, which displays a very small raw increase from $z\sim1$ to
$z\sim0$ for hosts with $M_\ast\lesssim10^{11}$ $h^{-2} M_\Sun$.
Below, the intrinsic growth in $M_{200}/M_\ast$ with time will refer
to growth after removing the factor of $1.3$.

To facilitate comparisons between $z\sim1$ and $z\sim0$, in the left
hand panels of Figure \ref{f:ml_data} we additionally plot the
satellite velocity dispersion and $M_{200}/L$ for galaxies at $z\sim1$
where all $M_B$ values have been dimmed by one magnitude to account
for evolution in $M^\ast$ between the median redshifts of the two
samples \citep[$=1.37\times(0.84-0.06)$; cf.][]{Faber05}.  This brings
the two samples into near agreement.  Our results as a function of
$M_B$ are thus consistent with host galaxies at $z\sim1$
\emph{evolving into} host galaxies at $z\sim0$ if their luminosities
on average dim by one magnitude and their dark matter halo masses do
not grow between these epochs.  However, we emphasize that since this
interpretation neglects the growth of dark matter halos with time it
is rather unrealistic, and return to a more detailed interpretation of
these results in $\S$\ref{s:disc2}.

\subsection{Results as a Function of Host Galaxy Color}

Finally, we investigate the $\sigma-L_B$ relation as a function of
host galaxy $U-B$ color.  The $\sigma-M_\ast$ relation cannot be
probed as a function of color because the high/low stellar mass bins
are almost entirely dominated by red/blue galaxies, so it is
impossible to separate the host population into red and blue at these
extremes of the stellar mass distribution and obtain adequate
statistics, but these effects are more modest for $\sigma$ vs. $M_B$.
Figure \ref{f:ml_color} and Table \ref{t:res1} present the satellite
velocity dispersion as a function of host galaxy luminosity, color, and
redshift.  We can only probe one broad bin in host luminosity ($-20.5<
M_B-5\rm{log}$$(h) <-22.0$) at $z\sim1$ due to the limited number of
satellites available.  For every bin in host $B$-band absolute
magnitude there is a clear dependence of satellite velocity dispersion
on host $U-B$ color, at both $z\sim1$ and $z\sim0$.

\section{Comparison to a Semi-Analytic Model}\label{s:sam} 

We now compare our results to predictions from a semi-analytic model
(SAM) of galaxy formation based on the Millennium Run (MR) $N$-body
simulation \citep{Springel05}, which is described in detail in
\citet{Croton06}.  In brief, the model evolves the contents of dark
matter halos self-consistently according to a set of simple physical
prescriptions derived from observational and theoretical phenomenology
that govern the evolution of baryons in a cosmological setting. These
prescriptions track a wide range of physics, including the cooling of
baryons, star-formation in galactic disks and merger induced
starbursts, supernova and AGN feedback, the tidal stripping of gas,
and black hole growth.  This model matches an array of observational
results at $z\sim0$, including the $b_j-$ and $K-$band galaxy
luminosity functions and the color bimodality visible in the
color-magnitude relation.

In order to a facilitate direct comparison to our observational
results, mock surveys have been constructed out of the MR that match
the geometry and sampling rates of both the DEEP2 and SDSS surveys.
We then apply exactly the same search criteria and analysis methods to
these mock surveys as to the data.

\begin{figure}
\centering
\plotone{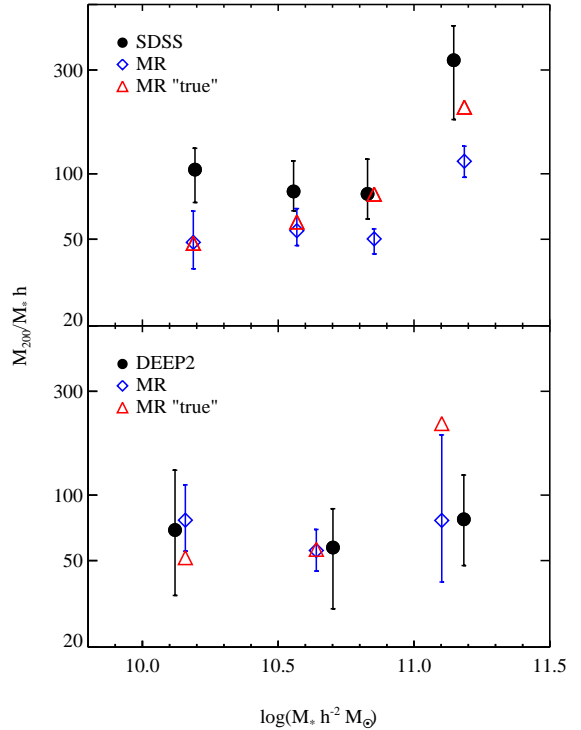}
\vspace{0.5cm}
\caption{Observed and predicted virial-to-stellar mass ratios,
  $M_{200}/M_\ast$, at $z\sim0$ (\emph{top panel}), and $z\sim1$
  (\emph{bottom panel}).  The observed virial-to-stellar mass ratios
  (\emph{circles}) are compared to both the true average
  virial-to-stellar mass ratio (\emph{triangles}) in the MR
  semi-analytic model and the ratio derived from applying our
  satellite kinematics measurement techniques to that sample
  (\emph{diamonds}).}
\vspace{0.5cm}
\label{f:sams}
\end{figure}

Figure \ref{f:sams} presents a comparison of $M_{200}/M_\ast$ between
the MR SAM and data at $z\sim0$ (\emph{top panel}) and $z\sim1$
(\emph{bottom panel})\footnote{The MR galaxy formation model was run
  with a Salpeter IMF; in order to compare to the observational
  stellar masses, which were computed with a Chabrier IMF, the MR
  stellar masses have been lowered by the known offset between these
  IMFs, $0.3$dex.}.  The $M_{200}/M_\ast-M_\ast$ relations for MR host
galaxies have shapes quite similar to the observed relations at both
low and high redshift (note that the highest stellar mass bin at
$z\sim1$ in the MR has only $13$ satellites and hence that measurement
is unstable).  In addition, the normalization of this relation
in the MR is in agreement with observations at $z\sim1$, though
systematically lower compared to observations at $z\sim0$.  This
qualitative agreement highlights the power of current generation
N-body simulations, including the MR, which explicitly follow
sub-structure within each virialized dark matter halo.  Such
sub-structures host the population of satellite galaxies from which
our analysis is derived, and are clearly important for accurate
modeling of the internal dynamics of group and cluster systems.

The MR also provides an explicit test of our methodology since we can
compare the halo mass inferred from satellite kinematics to the true
average host dark matter halo mass, which is readily available from
the MR.  The good agreement in Figure \ref{f:sams} between the true
and satellite-derived masses is very encouraging, though not
unexpected based on previous tests of the satellite methodology
(cf. Appendix \ref{s:appraisal}).  The agreement is less than ideal at
the highest stellar mass bins at both $z\sim1$ and $z\sim0$, though at
$z\sim1$ the discrepancy is of limited significance due to the modest
number of pairs in the MR in that bin.

At $z\sim0$ the discrepancy between true and satellite-derived halo
masses is potentially more interesting.  At this late epoch the MR SAM
contains many satellites whose host subhalos have been stripped below
the resolution limit of the simulation (the so-called ``orphan''
population).  In this situation the MR implements a prescription for
the dynamical evolution of such galaxies rather than tracking its
evolution explicitly with the subhalo.  When such galaxies are removed
from the MR, the agreement between true and satellite-derived masses
in the two highest stellar mass bins at $z\sim0$ significantly
improves while simultaneously preserving the agreement at lower
stellar masses and higher redshift.  This issue clearly warrants
further study.

\section{Discussion}\label{s:disc}

\subsection{Compatibility with Weak Lensing Results}

Our results at $z\sim0$ are most directly comparable to the weak
lensing results of \citet{Hoekstra05} who measured the lensing signal
around a sample of isolated galaxies observed by the Red-Sequence
Cluster Survey.  They found that $M_{\rm{vir}}/L\propto L^{0.5}$ for
isolated galaxies brighter than $M_B-5\rm{log}$$(h)=-20.0$, and
$M_{\rm{vir}}/L\sim$ constant for fainter galaxies.  Our results in
Figure \ref{f:ml_data} are consistent with these findings.  In
particular, note that our derived $M_{200}/L_B$ appears to flatten out
at $M_B\lesssim-20.0$, and we find a best-fit power law exponent of
$0.5$ for $M_{200}/L_B$ versus $L_B$ for the three data points
brighter than $M_B=-20.0$ (though taken together, our data are
consistent with a slope of $0.0$).  As discussed in
\citet{Hoekstra05}, their results are in broad agreement with other
weak lensing studies.  This overall consistency is quite encouraging
since these two methods for deriving virial masses are almost entirely
independent.

In addition there is rough agreement in the normalization of the
relations derived from the two techniques.  \citet{Hoekstra05} found
$M_{\rm{vir}}=1.9\times 10^{12}$ $h^{-1} M_\Sun$ for galaxies with
luminosity $M_B-5\rm{log}$$(h)=-20.6$ at $\bar{z}=0.32$.  In order to
compare our results to theirs, we have evolved our absolute magnitudes
by $1.37z$ to their median redshift, converted our definition of mass
to theirs (their definition of virial mass is a region that encloses a
mean density $\Delta_{\rm{vir}}$ times the mean density of the
universe, where $\Delta_{\rm{vir}}=337$ at $z=0$ and varies with
redshift), and corrected for the $\sim30$\% over-prediction in our
derived mass due to incompleteness (cf. Appendix \ref{s:contam}).
With these corrections we find a virial mass of
$M_{\rm{vir}}=1.8\times10^{12}$ $h^{-1} M_\Sun$ for host galaxies with
a mean absolute magnitude $M_B-5\rm{log}$$(h)=-20.4$.  While this
agreement is encouraging, at lower luminosities our corrected masses
tend to be higher than those quoted in \citet{Hoekstra05}, by about
$40$\% (though they still agree within $2\sigma$).

\citet[][M05]{Mandelbaum06} measured the average virial-to-stellar
mass ratio for galaxies in the SDSS as a function of galaxy stellar
mass and found that $M_{200}/M_\ast=37\pm13$ $h$ ($47\pm14$ $h$) for
early-type galaxies with $\rm{log}(M_\ast)=10.5 (10.7)$ $h^{-2}
M_\Sun$ (confidence levels quoted in M05 are $95$\%).  In the present
work we find $M_{200}/M_\ast=64^{+25}_{-12} $ $h$ ($62^{+28}_{-15}$
$h$) for host galaxies with $\rm{log}(M_\ast)=10.6 (10.8)$ $h^{-2}
M_\Sun$, at $z\sim0$ (after correcting for the $\sim30$\% overestimate
in our masses due to incompleteness; cf. Appendix \ref{s:contam}).  Our
results are hence quite consistent with M05 at the $95$\% confidence
level.  In addition, including the stellar mass of satellites results
in a virial-to-stellar mass ratio of the entire system for
$(M_{200}/M_\ast)_{\rm{tot}}=52^{+20}_{-9} $ $h$ for host-satellites
systems with $\rm{log}(M_{\ast,{tot}})=10.6$ $h^{-2} M_\Sun$.

Perhaps most importantly, our results agree with M05 in the highest
stellar mass bin.  For host galaxies with $\rm{log}(M_\ast)=11.1$
$h^{-2} M_\Sun$ we find $M_{200}/M_\ast=256^{+122}_{-119}$ $h$ (again
corrected for the $\sim30$\% overestimate in $M_{200}$) while M05
found $M_{200}/M_\ast=284^{+49}_{-75}$ $h$ for early-type galaxies
with mean stellar mass $\rm{log}(M_\ast)=11.3$ $h^{-2} M_\Sun$.

Most recently, \citet[][H06]{Heymans06} measured the weak lensing
signal for galaxies in the COMBO-17 and GEMS surveys.  These authors
found that the virial-to-stellar mass ratio decreases from $z\sim0.8$
to $z\sim0$ by at most a factor of $2.6$ ($1\sigma$ confidence) and
are consistent with a constant value of
$M_{200}/M_\ast=66^{+15}_{-20}$ $h$ over this redshift range.  These
authors include only galaxies with $\mathrm{log}(M_\ast) >
10.2$ $h^{-2} M_\Sun$, where there data is complete for $z<0.8$.  The
evolution of $M_{200}/M_\ast$ measured in the present work is again in
good agreement with these weak lensing results, including the
non-evolving virial-to-stellar mass ratio from $z\sim1$ to $z\sim0$
for galaxies with $M_\ast \lesssim 10^{11}$ $h^{-2} M_\Sun$ (the
average stellar mass of galaxies in H06 is $\langle
\mathrm{log}(M_\ast) \rangle =10.5$ $h^{-2} M_\Sun$).

A more detailed comparison between these two methods for estimating
halo masses is complicated because the systems that satellite dynamics
probe are systematically different from the systems weak-lensing
analyses explore.  In order to uncover any measurement bias one would
need to measure the weak-lensing signal for the same set of galaxies
that are used in satellite dynamics studies, which is likely feasible
given the current number of host galaxies found in the SDSS.

\subsection{The Evolution of Host Galaxies from $z\sim1$ to
  $z\sim0$}\label{s:disc2}

We now explore the evolution in the virial mass-to-light and
virial-to-stellar mass ratios -- i.e., $M_{200}/L_B$ and
$M_{200}/M_\ast$, respectively -- between $z\sim1$ and
$z\sim0$.  Evolution in $M_{200}/M_\ast$ is somewhat simpler to
interpret because stellar mass can only increase with time, and so we
consider it first.

Our data suggest a \emph{bimodality} in the evolutionary history of
host galaxies between $z\sim1$ and $z\sim0$: host galaxies below
$M_\ast\sim 10^{11}$ $h^{-2} M_\Sun$ maintain a roughly constant
virial-to-stellar mass ratio (the intrinsic ratio increases by a
factor of $1.1\pm0.7$) while hosts above this stellar mass scale,
which are predominantly red in $U-B$ color, experience a factor of
$4\pm3$ increase in $M_{200}/M_\ast$.  The quenching of star
formation, which becomes particularly effective above this stellar
mass \citep{Dekel06, Croton06, Cattaneo06}, is a key
physical process likely responsible for this bimodality.

Galaxies below $M_\ast\sim 10^{11}$ $h^{-2} M_\Sun$ continue to form
stars between $z\sim1$ and $z\sim0$.  Recent modeling of the star
formation history of relatively low-mass blue galaxies has suggested
that these objects grow in stellar mass by roughly a factor of $2$
between these epochs \citep{Noeske06a}.  Cosmological simulations of
dark matter reveal that $\sim10^{12}$ $h^{-1} M_\Sun$ dark matter
halos -- such as the ones which our typical host galaxies appear to
reside in -- grow on average by a factor of $\sim2$ over this time
period \citep{Wechsler02}.  The growth in stellar mass of host
galaxies hence proceeds in lockstep with the growth of their dark
matter halos, yielding a constant virial-to-stellar mass ratio from
$z\sim1$ to $z\sim0$.

If the virial-to-stellar mass ratio for these galaxies does not change
with $z$, their star formation efficiency must also remain constant
over this time.  Specifically, after correcting for the $\sim30$\%
overestimate in our masses due to incompleteness (cf. Appendix
\ref{s:contam}), our results imply a star formation efficiency for
host galaxies of $\eta\equiv \frac{M_\ast}{M_{200}}
\frac{\Omega_m}{\Omega_b} \approx0.13$, assuming a universal baryon
fraction of $0.17$ \citep{Spergel06} and $h=0.7$.  Including the
stellar mass of the satellite galaxies yields an efficiency for the
entire system of $\eta\approx0.16$.  These efficiency factors are
bracketed by values of $\eta=0.08$ derived from the global baryonic
mass function \citep{Read05} and various weak-lensing studies, which
find $\eta\approx0.05-0.3$ \citep{Mandelbaum06,Heymans06}.

The lack of evolution in $M_{200}/M_\ast$ found here is broadly
consistent with measurements of the evolution of the stellar mass
Tully-Fisher relation (TFR) over this time period.  In particular,
\citet{Conselice05} and \citet{Boehm06} converted their stellar mass
TFR into a dynamical-to-stellar mass ratio and found little evolution
from $z\sim1$ to $z\sim0$; using $\sim550$ galaxies, \citet{Kassin06b}
found no evolution in the stellar mass TFR between these epochs, which
under simple assumptions translates into a non-evolving
dynamical-to-stellar mass ratio.  One should keep in mind that there
are numerous details which make a one-to-one comparison between the
stellar mass TFR and our results complicated.  For example, the
relation between internal galaxy kinematics (e.g. the maximum circular
velocity of the disk) and dark matter halo kinematics is non-trivial,
and likely changes with time.  Recent SPH simulations of individual
disk galaxies have found that stellar mass and circular velocity
co-evolve (both roughly doubling between $z\sim1$ and $z\sim0$),
yielding a non-evolving stellar mass TFR \citep{Portinari06}.

The evolutionary history of the most massive ($M_\ast\gtrsim 10^{11}$
$h^{-2} M_\Sun$, and therefore red) host galaxies is predicted to be
qualitatively different.  A variety of recent models suggest that
feedback mechanisms in high mass halos prevent the ambient gas from
cooling and forming new stars \citep{Cattaneo06, Dekel06, Croton06}.
Hence the stellar mass of these galaxies can only increase via merging
with other galaxies.  Numerous observational results suggest that the
stellar mass in bright red galaxies \emph{at most} doubles since
$z\sim1$ \citep{Borch06, Bundy05, Bell04}.  Yet massive dark matter
halos are growing rapidly between $z\sim1$ and $z\sim0$
\citep{Wechsler02}.  The quenching of star formation within these dark
matter halos provides a natural, qualitative, explanation for the
observed growth in $M_{200}/M_\ast$ with time.

The quenching of star formation in these massive systems implies that
their star formation efficiency decreases with time since new gas that
comes into these halos is prevented from condensing to form new stars.
The observed increase in $M_{200}/M_\ast$ between $z\sim1$ and
$z\sim0$ for these systems corresponds to a decrease in efficiency
from $14$\% to $3$\% between these epochs.

This simple picture is complicated by our selection criteria.  In
particular, host galaxies are required to not have any comparably
bright companions within a search cylinder.  This non-trivial
selection makes it difficult to reason generally about the growth of
halos and stellar mass with time, and favors comparisons to more
detailed models of galaxy formation (see $\S$\ref{s:sam} and below).
However, these selection effects may also account for part of the
observed increase in $M_{200}/M_\ast$ between $z\sim1$ and $z\sim0$.

Due to the continual merging of galaxies within halos over time,
galaxies can enter the host galaxy sample between $z\sim1$ and
$z\sim0$.  In particular, objects within groups and clusters of
galaxies at $z\sim1$, which are in halos that are more massive than
the average host galaxy dark matter halo, may not be classified as
host galaxies due to their bright companions, but can \emph{turn into}
what we would call host galaxies at late times.  As groups and
clusters at $z\sim1$ evolve, dynamical friction will cause some
fraction of bright companion galaxies to merge with the central
galaxy.  Those systems in which only one luminous galaxy remains
within a massive dark matter halo will be similar to so-called
``fossil groups'' \citep[see e.g.][]{Jones03}.  The increase of fossil
groups between $z\sim1$ and $z\sim0$ is a possible explanation for the
increase in $M_{200}/M_\ast$ for high-stellar-mass host galaxies.
This mechanism will likely not affect the evolution of
$M_{200}/M_\ast$ at smaller host stellar masses because the merging of
$\sim L^\ast$ galaxies in massive halos will generate massive red
galaxies \citep{Hopkins06a}, not lower-mass, blue galaxies.

The stellar mass and redshift dependence of the virial-to-stellar mass
ratio of host galaxies is reproduced qualitatively in the Millennium
Run semi-analytic model \citep{Croton06}.  In particular, this model
manages to capture the stellar mass dependence of $M_{200}/M_\ast$ at
$z\sim1$ quite well.  At $z\sim0$ this model reproduces the observed
\emph{shape} of the $M_{200}/M_\ast-M_\ast$ relation, though with a
lower normalization.  The Millennium Run, and models like it, are
particularly useful for interpreting these results because these
semi-analytic models track the \emph{evolutionary history} of mock
galaxies.  Hence, for example, we are capable of directly confronting
the explanations presented above with a self-consistent model.  We can
ask where host galaxies at $z\sim1$ end up at $z\sim0$ and,
conversely, where $z\sim0$ host galaxies were at $z\sim1$.  Such a
comparison will be the focus of future work.

Evolution in the relation between host galaxy virial mass and $B$-band
absolute magnitude is more difficult to interpret because $M_B$ can
both increase, due to recent episodes of star-formation and mergers,
and decrease, due to fading in the stellar population.  Although it
was demonstrated earlier that our results as a function of $M_B$ are
consistent with the average host galaxy $M_B$ fading by one magnitude
from $z\sim1$ and $z\sim0$ while its dark matter halo does not evolve,
our results as a function of stellar mass cast this simple
interpretation into doubt.  In particular, evolution in the
$M_{200}/M_\ast$ relation strongly suggests that the average halo mass
of host galaxies is increasing from $z\sim1$ to $z\sim0$, since the
stellar mass of host galaxies is most likely increasing with time
(and, moreover, all currently favored cosmological models indicate
that halos grow in mass with time).  If halo mass increases
significantly, host galaxies at $z\sim1$ must fade by less than one
magnitude (i.e. fading by less than $L^\ast$ between these epochs) in
order to reach the $z\sim0$ $M_{200}/L_B$ relation (cf. Figure
\ref{f:ml_data}, left panels).

\section{Summary}\label{s:conc}

We summarize our main results and conclusions:

\begin{itemize}

\item[1.]  The $U-B$ color distribution of satellite host galaxies is
  indistinguishable from the colors of all galaxies (of similar
  luminosities), at both $z\sim1$ and $z\sim0$.  Satellites of host
  galaxies at $z\sim1$ are on average slightly \emph{fainter} in the
  $B$-band relative to their host galaxy compared to $z\sim0$, and
  host galaxies on average have a comparable number of satellites at
  both epochs.

\item[2.]  The line-of-sight velocity dispersion, $\sigma$, of
  satellites increases with the average host galaxy luminosity, $L_B$,
  and stellar mass, $M_\ast$.  In particular, $\sigma\propto
  M_\ast^{0.4\pm0.1}$ at both $z\sim1$ and $z\sim0$, while
  $\sigma\propto L_B^{0.6\pm0.1}$ at $z\sim1$ and $\sigma \propto
  L_B^{0.4\pm0.1}$ at $z\sim0$ (though it should be noted that
  these exponents are sensitive to the precise range of $L_B$ and
  $M_\ast$ over which the relation is fit).  In addition, at fixed
  $M_B$ red host galaxies have larger satellite velocity dispersions 
  compared to blue hosts, both at $z\sim1$ and $z\sim0$.

\item[3.]  The virial-to-stellar mass ratio, $M_{200}/M_\ast$, for
  host galaxies with $M_\ast\lesssim10^{11}$ $h^{-2} M_\Sun$ increases
  by a factor of $1.1\pm0.7$ from $z\sim1$ to $z\sim0$; the ratio does
  not evolve significantly between these epochs.  Host galaxies with
  stellar mass above $M_\ast\sim10^{11}$ $h^{-2} M_\Sun$, which are
  predominantly red in restframe $U-B$ color, experience an increase
  in $M_{200}/M_\ast$ of a factor of $4\pm3$ between these epochs.

\item[4.] The Millennium Run semi-analytic model of galaxy evolution
  reproduces the observed trends of $M_{200}/M_\ast$ as a function of
  $M_\ast$ at $z\sim1$ quite well, while at $z\sim0$ the model
  reproduces the shape of the $M_{200}/M_\ast-M_\ast$ but with a
  somewhat lower normalization compared to observations.

\end{itemize}

\acknowledgments C.C. and F.P. acknowledge the ``Palace Camp'', where
none of this work took place.  C.C. thanks Andrey Kravtsov, Simon
White and Chris Wolf for useful discussions and Katherine Malinowska
for a detailed reading of the manuscript.  The DEIMOS spectrograph was
funded by NSF grant ARI92-14621 and by generous grants from the
California Association for Research in Astronomy, and from UCO/Lick
Observatory. The DEEP2 survey was founded under the auspices of the
NSF Center for Particle Astrophysics.  The bulk of this work was
supported by National Science Foundation grants AST 95-29098 and
00-71198 to UCSC and AST 00-71048 to UCB.  Additional support came
from NASA grants AR-05801.01, AR-06402.01, and AR-07532.01 from the
Space Telescope Science Institute, which is operated by AURA, Inc.,
under NASA contract NAS 5-26555.  C.C. and F.P. acknowledge support
from the Spanish MEC under grant PNAYA 2005-07789.  A.L.C. and
J.A.N. are supported by NASA through Hubble Fellowship grants
HF-01182.01-A and HST-HF-01165.01-A, awarded by the Space Telescope
Science Institute, which is operated by the Association of
Universities for Research in Astronomy, Inc., for NASA, under contract
NAS 5-26555.  Some of the data presented herein were obtained at the
W.M. Keck Observatory, which is operated as a scientific partnership
among the California Institute of Technology, the University of
California and the National Aeronautics and Space Administration. The
Observatory was made possible by the generous financial support of the
W.M. Keck Foundation.  We also wish to recognize and acknowledge the
highly significant cultural role and reverence that the summit of
Mauna Kea has always had within the indigenous Hawaiian community; it
is a privilege to be given the opportunity to conduct observations
from this mountain.

Funding for the Sloan Digital Sky Survey (SDSS) has been provided by
the Alfred P. Sloan Foundation, the Participating Institutions, the
National Aeronautics and Space Administration, the National Science
Foundation, the U.S. Department of Energy, the Japanese
Monbukagakusho, and the Max Planck Society. The SDSS Web site is
http://www.sdss.org/.

The SDSS is managed by the Astrophysical Research Consortium (ARC) for
the Participating Institutions. The Participating Institutions are The
University of Chicago, Fermilab, the Institute for Advanced Study, the
Japan Participation Group, The Johns Hopkins University, Los Alamos
National Laboratory, the Max-Planck-Institute for Astronomy (MPIA),
the Max-Planck-Institute for Astrophysics (MPA), New Mexico State
University, University of Pittsburgh, Princeton University, the United
States Naval Observatory, and the University of Washington.

This work made extensive use of the NASA
Astrophysics Data System and of the {\tt astro-ph} preprint archive at
{\tt arXiv.org}.


\begin{appendix}

  \section{Impact of survey completeness on the derived average host
    galaxy halo mass}\label{s:contam}

\begin{figure}
\centering
\plotone{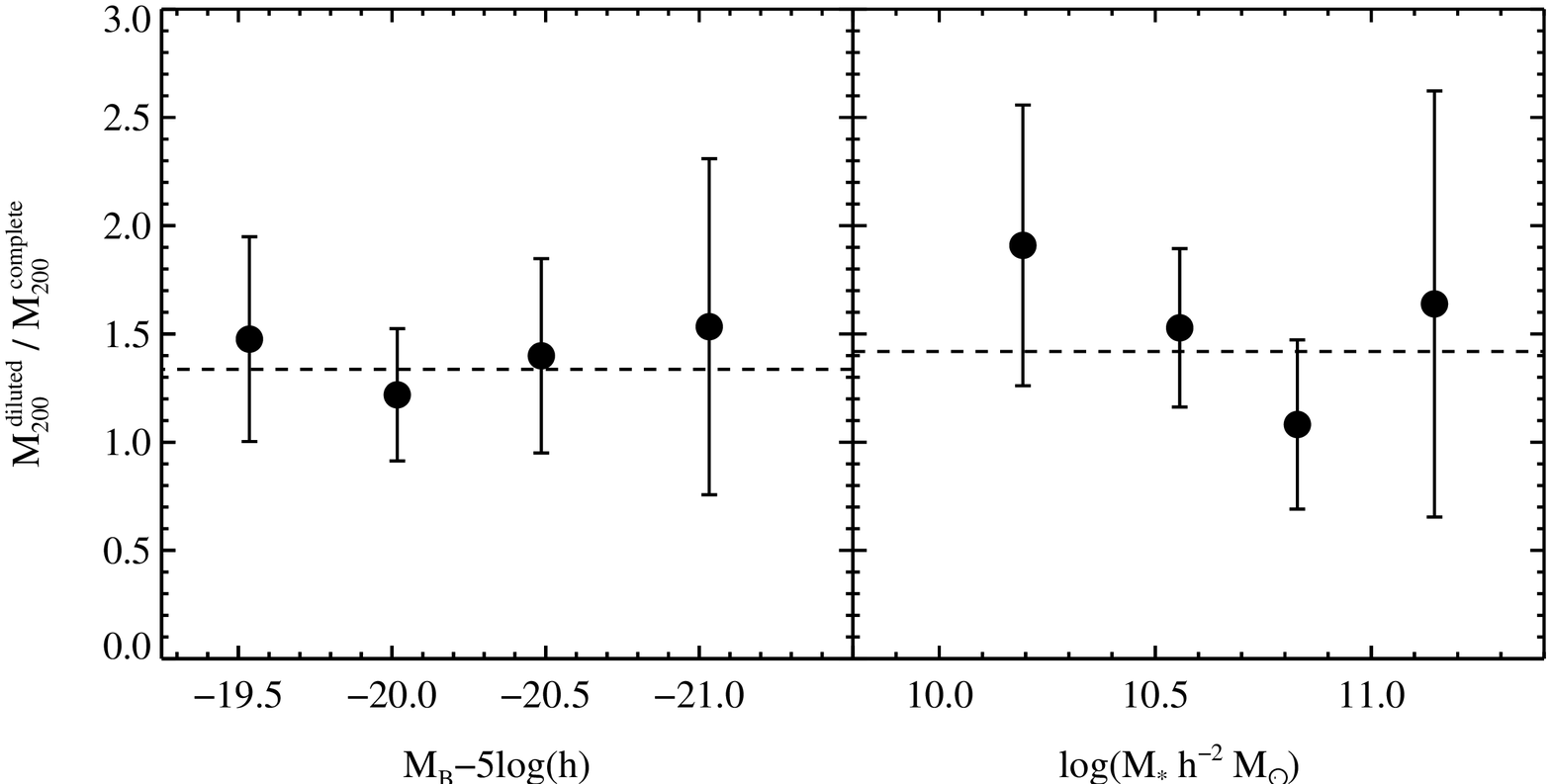}
\vspace{1.0cm}
\caption{Ratio of host galaxy halo masses derived from the complete
  SDSS survey to masses derived from the SDSS survey diluted by $40$\%
  to match the completeness of the DEEP2 survey.  The
  diluted sample yields systematically higher virial masses as a
  function of both $M_B$ (\emph{left panel}) and $M_\ast$ (\emph{right
    panel}).  The dashed line in each panel represents the average
  $M_{200}^{\mathrm{diluted}}/M_{200}^{\mathrm{complete}}$ value.  See
  Appendix \ref{s:contam} for details.}
\vspace{0.5cm}
\label{f:ml_fx}
\end{figure}

As described in $\S$\ref{s:deep2}, the DEEP2 survey obtains redshifts
for $\sim 50$\% of all galaxies with $R<24.1$.  In an incomplete
galaxy catalog, we may identify a galaxy as isolated when in reality
it has a bright companion that simply failed to be targeted.  We
assess the impact of incompleteness on the recovered velocity
dispersion and halo mass by randomly diluting the SDSS survey and then
searching for host galaxies and satellites.  We show in Figure
\ref{f:ml_fx} the change in the measured halo mass, $M_{200}$, as a
function of $M_B$ and $M_\ast$, comparing host and satellite galaxies
identified in the complete SDSS to the diluted SDSS.  The recovered
masses agree within $\sim 1\sigma$, which is encouraging, as this
demonstrates that the modest incompleteness of the DEEP2 survey does
not greatly change the recovered dispersion.

There is, however, a noticeable bias in the recovered masses: the
diluted samples yield higher masses at a given host galaxy $M_B$ and
$M_\ast$.  This is due to the inclusion of systems that are not truly
isolated, but in fact reside in denser environments, which will have
on average larger masses than isolated systems.  We find that
$\sim35$\% of host galaxies identified in the diluted sample are in
fact not truly isolated.  This number is comparable to the
contamination rate found in \citet{Conroy05a}, who analyzed mock
catalogs that included the same selection effects as the DEEP2 survey.
This contamination fraction depends mildly on host luminosity, with
sub-samples of brighter hosts having a lower contamination fraction
than sub-samples of fainter hosts.  This trend reflects the fact that
brighter galaxies tend to be found in denser environments, so they are
more likely to still have bright companions after dilution.  The
numerous bright companions will help prevent a bright galaxy from
being falsely identified as isolated in the diluted sample, while the
relative paucity of similar luminosity companions around fainter
galaxies will make the identification of such a galaxy as (falsely)
isolated in a diluted sample easier.  

The average ratio between diluted and complete SDSS host galaxy halo
masses is indicted by the dashed lines in Figure \ref{f:ml_fx}.  We
use these average mass corrections to arrive at host galaxy halo
masses for $100$\% complete surveys, both at $z\sim0$ and $z\sim1$,
when comparing against other work.

These falsely isolated host galaxies probably do not strongly effect
the recovered velocity dispersion because the satellites that they
contribute to the $dV$ distribution resemble the interloper
population: they are a small fraction of the total sample and they are
more uniformly distributed in $dV$ compared to satellites from
truly isolated host galaxies.

Finally, we point out that any potential bias that may result from
incompleteness will only impact the \emph{absolute} $\sigma$ and
$M_{200}/L$ values quoted herein.  Results concerning the
\emph{relative} change in these quantities between $z\sim1$ and
$z\sim0$ are likely not affected by incompleteness, since we have
diluted the SDSS down to the same completeness as DEEP2 for those
studies.

\section{Impact of changing measurement parameters}\label{s:test_mass}

\begin{deluxetable*}{c|cc|ccc|ccc|ccc}
\tablecaption{Dependence of derived virial mass on various parameters} 
 \tablehead{
  \multicolumn{1}{c}{Sample} &
  \multicolumn{2}{|c}{Anisotropy} & 
  \multicolumn{3}{|c}{Concentration } &
  \multicolumn{3}{|c}{Search Criteria } &
  \multicolumn{3}{|c}{Radial bins ($h^{-1}$ kpc)} \\
  \colhead{} & 
  \colhead{$\beta=0$} &
  \colhead{$\beta_1$} & 
  \colhead{$c=5$} & 
  \colhead{$c=10$} & 
  \colhead{$c=15$} & 
  \colhead{$A$} & 
  \colhead{$B$} & 
  \colhead{$C$} &
  \colhead{$[20,100]$} & 
  \colhead{$[20,150]$} & 
  \colhead{$[20,200]$} 
\\
 }
\startdata
SDSS & $2.8\pm0.6$ & $2.8\pm0.6$ & $2.8\pm0.7$ & 
$2.8\pm0.6 $& $2.8\pm0.6$ &  $2.8\pm0.6$ & $3.0\pm0.8$ & 
$3.0\pm1.0$ & $2.3\pm0.6$  & $2.8\pm0.6$ & $3.5\pm0.7$ \\
DEEP2 & $1.5\pm0.6$ & $1.6\pm0.7$ & $1.3\pm0.6$ & 
$1.5\pm0.6$ & $1.6\pm0.6$ & $1.5\pm0.6$ & $1.3\pm1.0$ &
$2.5\pm2.5$ & $1.0\pm0.5$ & $1.5\pm0.6$ & $2.5\pm1.0$ \\
\enddata
\tablecomments{The SDSS and DEEP2 samples include host galaxies with
  $-20.25<M_B-5\rm{log}$$(h)<-19.75$ and $-21.0<M_B-5\rm{log}$$(h)<-19.5$,
  respectively.  All values are virial masses ($M_{200}$) in units of
  $10^{12}$ $h^{-1} M_\Sun$.  The fiducial set of assumptions and
  parameters are $\beta=0$, $c=10$, search criteria $A$, and radial
 bin $[20,150]$ $h^{-1}$ kpc.  Each table entry is the result of
  varying only one of these parameters at a time.  See Table
  \ref{t:params} for details regarding the search criteria. \vspace{0.5cm}}
\label{t:tt}
\end{deluxetable*}

In this section we explore the effects of varying the parameters we
have used when measuring velocity dispersions and deriving virial
masses.  Specifically, we explore the impact of changing the
host-satellite search criteria or the number and size of the radial
bins in which we measure velocity dispersions, and in addition we show
that the derived masses are insensitive to the assumed halo
concentration and velocity anisotropy.  For simplicity, in the following
tests we have selected one sample at $z\sim0$ and one at $z\sim1$ to
illustrate our results (see Table \ref{t:tt}), and the $z\sim0$ sample
has been diluted to match the completeness of the sample at $z\sim1$,
as mentioned previously.  Our fiducial set of assumptions and
parameters are an isotropic velocity distribution ($\beta=0$), an NFW
concentration of $c=10$, a bin in projected separation between
satellite and host of $R_p=[20,150]$ $h^{-1}$ kpc, and search criteria
$A$.

We begin by testing assumptions that affect the measurement of
velocity dispersions (and through this, the derived masses).  In Table
\ref{t:tt} we show the change in the derived virial mass when we use
different search criteria and different radial bins.  The primary
effect of changing search criteria is that the number of satellites
changes, which in turn alters the errors on the recovered velocity
dispersion and virial mass.  For example, using search criteria $B$ or
$C$ at $z\sim1$ results in only $30$ and $32$ satellites within
$R_p=[20,150]$ $h^{-1}$ kpc, respectively, for the luminosity bins we
consider in Table \ref{t:tt}; hence the velocity dispersions and
masses estimated for these samples are rather unstable.  As can be
gleaned from the table, changing the search criteria or varying the
radial binning results in changes in the recovered masses within
$1\sigma$ of our fiducial estimates.  Note also that using search
criteria $B$ at $z\sim1$ and $C$ at $z\sim0$ amounts to using
consistent isolation criteria in \emph{comoving} coordinates (because
in this case the projected separation scales as $(1+z)^{-1}$), and
hence we conclude that using such coordinates, as opposed to physical
coordinates, results in no significant change in our results.

We now test the impact on $\sigma$ of using relatively bright
satellites, i.e satellites that are only one magnitude fainter than
their host galaxy.  One might expect more massive (brighter)
satellites to be out of virial equilibrium with the host galaxy since
dynamical friction acts more efficiently on these satellites compared
to less massive satellites.  Using the complete SDSS sample, we
measure the satellite velocity dispersion for hosts identified using
search criteria $A$ as a function of $dM$, the separation in
brightness between the satellite and host galaxies.  Specifically, we
only include satellites that are at least $1.0$, $1.5$, $2.0$, $2.5$,
and $3.0$ magnitudes fainter than their host galaxy, and find that the
dispersions estimated from these restricted samples all agree well
within $1\sigma$.

Since using different radial bins causes the largest changes in these
tests, we have explored its effects in more detail.  We have
reconstructed the $\sigma-L$ and $\sigma-M_\ast$ relations using each
set of radial bins and find no qualitative change in our results.  In
particular the exponents of these relations when using $R_p=[20,200]$
$h^{-1}$ kpc are the same at both $z\sim1$ and $z\sim0$, in agreement
with the results shown in Figure \ref{f:ml_data}.  The results for the
$R_p=[20,100]$ $h^{-1}$ kpc are noisier, as there are $\sim50$\% fewer
satellites in this sample, but are also consistent with our fiducial
results.

We now test assumptions that only affect the conversion between
velocity dispersion and mass.  In Table \ref{t:tt} we show the effects
of varying the concentration, $c$, or velocity anisotropy, $\beta$, on
the recovered virial mass.  We use concentrations of 5, 10, and 15,
which span the range of typical concentrations for $\sim 10^{12}$
$h^{-1} M_\Sun$ dark matter halos for $0<z<1$.  We test two forms
of the velocity anisotropy, the isotropic case ($\beta=0$), and a form
suggested by \citet{Mamon05} that is a good fit to a compilation of
anisotropies derived from clusters within $N$-body simulations:
\begin{equation}
\beta_1(r) =  \frac{1}{2}\frac{r}{r+r_a},
\end{equation}
where $r_a=0.18\,r_{200}$.  This functional form has not been tested
in galaxy-sized halos, though see the discussion in
Appendix \ref{s:appraisal}.

At first glance it seems somewhat remarkable that the derived virial
mass is almost independent of the concentration and velocity
anisotropy parameters over the range tested.  As it turns out, the
line-of-sight velocity dispersion profile in the regime $50<R_p<150 $
$h^{-1}$ kpc, which encompasses most of the satellites we use, is
almost completely insensitive to concentration and anisotropy, while
on both smaller and larger scales their effects are quite noticeable.
We are, by necessity if not luck, probing a ``sweet spot'' in the
velocity dispersion profile.

In addition, we have tested the sensitivity of the derived virial mass
to the number of radial bins used to measure velocity dispersions.  We
can only perform this test for the SDSS sample due to the limited size
of the DEEP2 sample.  We have measured the velocity dispersion in
three bins of projected separation, $R_p=[20-100], [100-200],$ and
$[200-300]$ $h^{-1}$ kpc, for the SDSS sample used in Table
\ref{t:tt}.  We find a best fit virial mass of $M_{200}=3.8\pm0.5
\times10^{12}$ $h^{-1} M_\Sun$, which is within $1\sigma$ of the
values quoted in Table \ref{t:tt} using only one radial bin.
Furthermore, rebinning the \emph{undiluted} SDSS sample results in a
mass of $M_{200}=2.6\pm0.6 \times10^{12}$ $h^{-1} M_\Sun$.  While we
see a decline in the velocity dispersion profile for the undiluted
sample in agreement with \citet{Prada03}, we see no decline for the
diluted sample.  This is likely due to the fact that the $35$\% of
host galaxies in the diluted sample that are not truly isolated
introduce noise into the dispersion measurement on large scales.

Finally, we turn to our assumption of an NFW \citep{NFW97}
characterization of the density profile.  Recently, \citet{Prada03}
have detected a decline in the satellite velocity dispersion profile
with increasing distance from the host galaxy at $z\sim0$.  This
decline distinctly favors an NFW density profile over an isothermal
distribution.  However, we cannot distinguish between these and other
density distributions at $z\sim1$ because there are an insufficient
number of satellites, making it difficult to bin more finely in radius
and probe the velocity dispersion \emph{profile}.

We are motivated to choose the same parameterization of the density
profile at $z\sim1$ and $z\sim0$ because it seems unphysical for the
density profile to drastically change, e.g. from isothermal to NFW,
over this interval.  Furthermore, an NFW-like density profile
(notwithstanding minor deviations) appears to be a generic feature of
hierarchical clustering, and is not sensitive to initial conditions or
cosmological parameters \citep[see e.g.][]{NFW97, Navarro04}.  Note
that our conclusions regarding the dependence of the satellite
velocity dispersion on host luminosity and redshift are completely
independent of density profile considerations.

\section{Assessment of Assumptions}\label{s:appraisal}

We identify three main assumptions inherent in using satellite
galaxies to probe the virial dark matter mass of their hosts: 1) there
is little scatter between host galaxy luminosity and dark matter halo
virial mass, 2) satellites are fair tracers of the underlying dark
matter velocity field, and 3) the velocity difference distribution of
satellites and interlopers can be modeled as a Gaussian and a
constant, respectively.  These assumptions are well motivated both
observationally and theoretically.

The first assumption (low scatter between mass and luminosity) must
hold for our stacking procedure to work; it is motivated by the
Tully-Fisher relation for disk-dominated galaxies and the
Faber-Jackson relation for bulge-dominated galaxies.  These relations
have only a modest amount of scatter; typically $0.2-0.9$ magnitudes
at fixed velocity.  The scatter is even smaller when these relations
are quoted as a function of stellar mass or total baryonic mass
\citep{Verheijen01, Mcgaugh05}.  These tight correlations, with the
assumption that the velocity measured probes the underlying halo mass,
suggest that the scatter between halo mass and galaxy luminosity is
not large, and hence that our stacking procedure is valid.  In
addition, models with a tight relation between galaxy luminosity and
halo mass successfully reproduce a wide variety of observations
\citep[see e.g.][]{Cole00, Tasitsiomi04, Conroy06a, Croton06, Vale04}.

Under the assumption that satellite galaxies can be identified with
subhalos in dissipationless $N$-body simulations, these simulations
support our second assumption, that satellites are fair tracers of the
underlying mass distribution.  A number of studies have looked for a
``velocity bias'' between dark matter and subhalos.  Recent work has
suggested that such a bias, defined as the ratio between the velocity
dispersion of subhalos and dark matter, is about $10$\% when averaged
over entire clusters \citep{Ghigna00,Diemand04}.  In fact, when
\citet{Faltenbacher06} identify subhalos in a simulated cluster in
such a way that their spatial distribution matches observed galaxies,
they find no velocity bias.  Although no study has systematically
investigated the velocity bias in galaxy-size halos, \citet{Prada03}
showed that subhalos identified in a simulated galaxy-sized halo
accurately reflect the underlying mass distribution.  Admittedly, the
velocity bias of the most massive subhalos, which should correspond to
the satellites studied here (since we use relatively bright
satellites, which are likely associated with the most massive
subhalos), has not been adequately investigated.  Nevertheless, we
conclude that satellite galaxies, should be fair tracers of the
underlying dark matter halo mass to $\sim10$\% or better.

The second and third assumptions have been tested in conjunction by a
number of authors.  \citet{Prada03} identified satellite galaxies with
subhalos in dissipationless $N$-body simulations, and artificially
added in interloper galaxies uniformly in phase space.  They found
that for an isolated Milky-Way sized dark matter halo, the Gaussian
plus constant parameterization accurately recovers the velocity
dispersion profile of subhalos, which in turn accurately reflects the
underlying dark matter halo mass.

\citet{VDB04b} also tested these assumptions using cosmological mock
galaxy catalogs at $z\sim0$, and found that the simple Gaussian plus
constant parameterization accurately recovers the velocity dispersion
of host galaxies as a function of galaxy luminosity.  These authors
additionally investigated the impact of orbital anisotropy and a
spatial (anti-) bias of the satellite galaxies, and found the
resulting effects on the recovered dispersion to be minimal.
Specifically, when using satellites within one-third of the virial
radius from the host galaxy, they found that if $\beta$ changes from
$-0.5$ to $0.5$ then the velocity dispersion changes by $<10$\%.

These assumptions have also been tested at $z\sim1$ by
\citet{Conroy05a} who used cosmological simulations into which mock
galaxies were inserted using a halo model approach \citep[see][for
details concerning these mock catalogs]{Yan03,Yan04}. These authors
found that the velocity dispersion profile of mock galaxies accurately
reflects the underlying halo mass, and that interlopers can be
effectively modeled as a constant component to the velocity difference
distribution.

Consistency between weak-lensing and satellite measurements of
$M_{200}/L$ in past studies gives further confirmation that satellites
are fair tracers of the underlying dark matter velocity field (unless
both satellite dynamics and weak lensing measurements are biased in a
similar way).  \citet{McKay02} compared the dependence of the
mass-to-light ratio on luminosity derived from both satellite dynamics
and weak lensing measurements, and found agreement within $1\sigma$.
Furthermore, both weak lensing studies \citep{Guzik02, Hoekstra04,
  Hoekstra05, Kleinheinrich05, Mandelbaum06} and satellite dynamics
\citep{McKay02, Prada03,Brainerd03, VDB04b} have found that the
derived virial mass scales with host luminosity as $M\propto L^\alpha$
with $1.0\lesssim\alpha\lesssim1.5$.  The range in $\alpha$ can be at
least partially explained by the different regimes (isolated galaxies
vs. groups and clusters) and photometric bands probed.

\end{appendix}

\end{document}